\documentclass[11pt]{article}%
\usepackage{amsfonts}
\usepackage{epsfig,amsmath,latexsym,amssymb}
\usepackage{amscd}
\usepackage{amsthm}
\usepackage{multirow}
\usepackage{graphicx}
\usepackage{geometry}
\usepackage{lscape}
\usepackage{amsmath}
\usepackage{amssymb}
\usepackage{bm}
\usepackage[ruled,vlined]{algorithm2e}
\providecommand{\U}[1]{\protect\rule{.1in}{.1in}}
\renewcommand{\baselinestretch}{1.5}

\newcommand{\Remm}[1]{}

\newtheorem{model ass}[theo]{Model Assumptions}

\numberwithin{equation}{section}
\renewcommand{\thesection}{\arabic{section}}
\begin{document}

\title{Calibration and filtering for multi factor commodity models with seasonality: incorporating panel data from futures contracts.}
\author{Gareth W.~Peters$^{1,2}$
\and Mark Briers$^{3}$ \and Pavel V.~Shevchenko$^{1,2}$ 
\and Arnaud Doucet$^{4}$
\\$^{[1]}${\small UNSW Mathematics and
Statistics Department, Sydney, 2052, Australia;}\\{\small email:
peterga@maths.unsw.edu.au; } \\$^{[2]}${\small CSIRO Mathematical
and Information Sciences, Sydney, Locked Bag 17, North
Ryde,}\\{\small NSW, 1670, Australia; e-mail:
Pavel.Shevchenko@csiro.au, fax: +61-2-93253200}
\\$^{[3]}${\small QinetiQ, UK.}
\\$^{[4]}${\small University of British Columbia, Department of Computer Science and Department of Statistics.}}

\date{\textbf{arXiv:} Working Paper, Version from \today }

\maketitle

\begin{abstract}
We examine a general multi-factor model for commodity spot prices and futures valuation. We extend the multi-factor long-short model in \cite{Schwartz98} and \cite{yan2002valuation} in two important aspects: firstly we allow for both the long and short term dynamic factors to be mean reverting incorporating stochastic volatility factors and secondly we develop an additive structural seasonality model. Then a Milstein discretized non-linear stochastic volatility state space representation for the model is developed which allows for futures and options contracts in the observation equation. We then develop numerical methodology based on an advanced Sequential Monte Carlo algorithm utilising Particle Markov chain Monte Carlo to perform calibration of the model jointly with the filtering of the latent processes for the long-short dynamics and volatility factors. In this regard we explore and develop a novel methodology based on an adaptive Rao-Blackwellised version of the Particle Markov chain Monte Carlo methodology. In doing this we deal accurately with the non-linearities in the state-space model which are therefore introduced into the filtering framework. We perform analysis on synthetic and real data for oil commodities.

\textbf{Key words: }Multi-Factor, Commodity, Spot Price,
Stochastic Volatility, Milstein, Adaptive Markov chain Monte Carlo, Particle filter, Rao-Blackwellization.
\end{abstract}

\section{Commodities economic theory}

The basis of commodity modelling has a long history in the economics and econometrics literature, indeed instrumental concepts were introduced in \cite{Keynes} which developed normal backwardation to attempt to model inter-temporal futures prices, in order to explain differences between spot and futures prices. Storage theory was developed in \cite{Working49} to account for storage costs, which basically states that inter-temporal prices are the jointly determined price of storage. That is, futures prices will be approximately the spot price plus positive storage costs. One problem with this explanation was that the interpretation of backwardation, which occurs when the observed Futures price is less than the Spot price, was difficult to explain. To address this problem with interpretation, the work of \cite{kaldor1939speculation} developed the concept of convenience yield, which is the implicit revenue associated with stock holding. This can be considered as analogous to coupons linked with bonds or dividends payed on stocks. Put another way the convenience yield can be interpreted as the flow of services accruing to the holder of the spot commodity but not to the owner of a futures contract. 

The conclusions of storage theory provided the following fundamental elements that models of commodities should aim to address when considering the economic relationship between spot and futures prices implied by storage cost. This relationship identifies at least three key factors influencing futures prices, given by the Spot price, the convenience yield and the interest rate. Additionally, convenience yield and spot price are positively correlated and are both inverse functions of stock level. One also needs to consider that there is an asymmetric relationship for basis, that is the variation between spot price of a deliverable commodity and the relative price of the futures contract closest to maturity. In particular when the market is in contango (futures $>$ spot), the basis level is limited to the storage costs. One final important economic relationship that commodity models should aim to address is the Samuelson Effect, \cite{samuelson1965proof}, which states that movement in prices of prompt (soon to mature) contracts are large and erratic whilst prices of long-term contracts are not erratic. This implies that variance of futures prices and correlation between nearest futures price and subsequent contract prices decrease as a function of time.

\subsection{Outline and Contributions}
In this paper we begin with a review of multi-factor commodity futures models. Then in Section 3 we propose and develop a general multi-factor model for commodity spot prices and futures valuation. We extend the multi-factor long-short model in \cite{Schwartz98} and \cite{yan2002valuation} in two important aspects: firstly we allow for both the long and short term dynamic factors to be mean reverting incorporating stochastic volatility factors and secondly we develop an additive structural seasonality model. In Section 4 a Milstein discretized non-linear stochastic volatility state space representation for the model is developed which allows for futures and options contracts in the observation equation. In Section 5 we develop an Bayesian filtering and calibration framework in which we develop numerical methodology based on an advanced Sequential Monte Carlo algorithm utilising Particle Markov chain Monte Carlo to perform calibration of the model jointly with the filtering of the latent processes for the long-short dynamics and volatility factors. Section 6 explains this novel methodology in detail. In this regard we explore and develop a novel methodology based on an adaptive Rao-Blackwellised version of the Particle Markov chain Monte Carlo methodology. In doing this we deal accurately with the non-linearities in the state-space model which are therefore introduced into the filtering framework. In Section 7 we perform analysis on synthetic and real data for oil commodities.

\section{Review of multi-factor dynamic commodity futures models}
The typical approach to formulating a model for commodities is to begin by specifying a real world latent stochastic differential equation (s.d.e.) based model for the factors to be considered in the commodities model. Then under assumptions on the market prices of risk and the behavior of such markets, a change of measure is made to adjust the real world process by the price of risk in order to obtain the risk neutral measure. Under this risk neutral s.d.e. model, the futures price are obtained as functions of the latent process factors as expected discounted future payoffs.

In \cite{brennan1985evaluating} a reduced form model for commodity prices was introduced. This was a one-factor model in which the commodity price follows a Geometric Brownian motion (GBM) and the convenience yield is treated as a dividend yield. This model was clearly not sufficient as it failed in two important aspects, firstly it does not account for the property of mean reversion of spot prices and secondly it neglects inventory dependence properties of the convenience yield. This basic model was extended by the work in \cite{gibson1990stochastic} to a two factor model and used to study oil futures. In the two-factor model the spot price was modelled by Geometric Brownian Motion and the Convenience Yield was modelled by a mean reverting process. 

As noted in \cite{Schwartz98}, this two factor model can be made more parsimonious via a change of variable. A new state variable is defined by the demeaned convenience yield minus the long-term convenience yield and at time $t$ this represents the short-term deviation in log prices or the deviation of spot price returns from its long-term value. Next, define a second factor as the long-term price return or price appreciation which is given by the long-term total return minus the long term convenience yield. This reformulation in addition to being parsimonious is also easier to interpret from a physical perspective since now the returns are defined in terms of the long-term price appreciation. 

In the three-factor model, (\cite{schwartz1997stochastic}; \cite{miltersen1998pricing}; \cite{hilliard1998valuation}) a third unobserved stochastic variable is introduced, with the short term variation (follows OU process), the equilibrium price (follows GBM process) and the instantaneous growth in equilibrium prices (follows a mean reverting process) for the real process.

Note, it should be pointed out that an additional assumption on these models is that there are no arbitrage opportunities. Strictly speaking this will not be true since, as pointed out in \cite{ribeiro2004two}, these models do not guarantee that the convenience yield is strictly positive. This is required for these models to be arbitrage free, where contemporaneous spot prices are greater than discounted futures prices net of carrying costs. Additionally, as pointed out in the introduction, a key element of storage theory is that the variance and correlation between spot and futures prices should be related to the level of the price or convenience yield. These models make the simplistic assumption of constant correlation and variance. Finally, in addition to these weaknesses, the spot price volatilities obtained from these models are strongly hetroskedastic, see \cite{ribeiro2004two} for a review of literature on studies demonstrating these issues.

More recent models include those developed in \cite{ribeiro2004two} which removes the potential for cash and carry arbitrage, by forcing the convenience yield to be positive at all times. Additionally, these new models admit time varying spot and convenience yield volatilities. This model modifies the models discussed whilst still admitting an analytic expression to be obtained for the futures contract price as a function of the unobserved stochastic factors. In particular the two factor model they develop has a spot price following a GBM in which the convenience yield is treated as an exogenous dividend yield and the volatility is proportional to the square root of the instantaneous convenience yield level. The second factor, the convenience yield, follows a Cox-Ingersoll-Ross (CIR) model, forcing convenience
yield to be strictly non-negative, with volatility proportional to the square root of the instantaneous convenience yield level. Under the assumption that the interest rate is constant, an analytic solution is obtained for the futures price as a function of these unobserved stochastic factors (real process).

Another recent model is from \cite{yan2002valuation} which developed a very general multi-factor model incorporating several extensions to the above models including a Levy driven process with jumps. However, again in this model the questions relating to joint parameter estimation and filtering were not addressed, nor was seasonality incorporated into the model. The stochastic model developed was given by (risk neutral setting):
\begin{align*}
&dS_t = \left(r_t - \delta_t - \lambda \mu_J\right)S_tdt + \sigma_S S_t dW_1 + \sqrt{V_t} S_t dW_2 + JS_tdq \\
&\text{Pr}(dq=1) = \lambda dt; \ln(1+J) \sim N\left(ln(1+\mu_J) - \frac{1}{2}\sigma^2_J,\sigma^2_J\right)\\
&dr_t = \left(\theta_r - \kappa_r r_t \right)dt + \sigma_r \sqrt{r_t}dW_r; d\delta_t = \left(\theta_{\delta} -\kappa_{\delta}\delta_t\right)dt + \sigma_{\delta}dW_{\delta}\\
&dV_t = \left(\theta_V - \kappa_{V} V_t\right)dt + \sigma_V
\sqrt{V_t}dW_V + J_{V}dq; J_V \sim \exp(\theta).
\end{align*}
Where $S_t$ is the spot price at time $t$, $\delta_t$ is the instantaneous convenience yield at time $t$, $\mu_J$ the long term total return (price appreciation + convenience yield), $\kappa_r$, $\kappa_V$ and $\kappa_{\delta}$ are the mean reverting coefficients, $\theta_r$ and $\theta_V$ are the long term interest rate and volatility, $\sigma_{s},\sigma_{r}$ the volatilities of spot price and convenience yield dynamics. 

We modify this model structure for this paper and derive the resulting futures price as a function of the latent commodity factors. To achieve this we remain in the exponentially affine family of models. The derivation of the futures price for our model allows us to use it as the observation equations for these models that we consider in this paper which is derived as a exponentially affine linear function of the state variables. The observation equations linking these unobserved factors to the futures price were obtained under the risk neutral measure.

\section{Proposed Multi-Factor Stochastic long-short price dynamics model incorporating seasonality.}
In this paper we will extend the long-short stochastic Model to allow both the long and short stochastic factors to be mean reverting processes and we incorporate an additive structural model to account for seasonality and a third factor for stochastic volatility to allow for modelling of implied volatility smiles.

\subsection{Latent stochastic factors and seasonality.}
Here we present the model considered in the remainder of this paper.
\begin{center}
\begin{tabular}
[l]{|ll|}\hline \textbf{Real World Process} & \textbf{Risk Neutral
Process}\\\hline 
\multicolumn{1}{|c|}{$X\left(t\right)=\ln\left(
S_{t}\right) =\chi_{t}+\xi_{t}+\theta_{t}+f(t)$} &
$X^{\ast}\left(t\right)  =\chi_{t}^{\ast}+\xi_{t}^{\ast}+\theta_{t}^{\ast}+f(t)$\\
\multicolumn{1}{|c|}{$d\chi_{t}=-\beta\chi_{t}dt+\sigma_{\chi}dZ_{\chi}$}
& $d\chi_{t}^{\ast}=\left(
-\beta^{\ast}\chi_{t}^{\ast}-\lambda_{0}\right)
dt+\sigma_{\chi}dZ_{\chi}^{\ast}$\\
\multicolumn{1}{|c|}{$d\xi_{t}=\left(\mu_{\xi} -
\kappa_{\xi}\xi_{t}\right)dt + \sigma_{\xi}dZ_{\xi}$} & $d\xi
_{t}^{\ast}=\left(\mu^{\ast}_{\xi} -
\kappa^{\ast}_{\xi}\xi^{\ast}_{t}\right)dt
 +\sigma_{\xi}dZ_{\xi}^{\ast}$\\
\multicolumn{1}{|c|}{$d\theta_{t}= \sqrt{V_t}dZ_{\theta}$} &
$d\theta_{t}^{\ast}= -\lambda_4V_t^{\ast}dt + \sqrt{V_t^{\ast}}dZ_{\theta}^{\ast}$\\
\multicolumn{1}{|c|}{$dV_{t}=\left(\mu_{V} -
\kappa_{V}V_{t}\right)dt + \sigma_{V}\sqrt{V_t}dZ_{V}$} &
$dV_{t}^{\ast}=\left(\mu_{V}^{\ast} -
\kappa_{V}^{\ast}V_{t}^{\ast}\right)dt +
\sigma_{V}\sqrt{V_t^{\ast}}dZ_{V}^{\ast},$\\\hline
\end{tabular}
\end{center}
with correlations structures given by:
\begin{align*}
&\text{E}\left[dZ_{\xi}dZ_{\theta}\right]=
\text{E}\left[dZ_{\chi}dZ_{\theta}\right]=
\text{E}\left[dZ_{\chi}dZ_{V}\right] =
\text{E}\left[dZ_{\xi}dZ_{V}\right] = 0;
\text{E}\left[dZ_{\xi}dZ_{\chi}\right] = \rho_{\xi\chi};
\text{E}\left[dZ_{V}dZ_{\theta}\right] = \rho_{V\theta}.
\end{align*}
In addition we define $f(t)$ as the deterministic unknown seasonal
component and the risk premia in the risk neutral model are given
by:
\begin{align*}
&\lambda_{\chi}\left(\chi_{t},\xi_{t},\theta_{t},V_t\right)=
\lambda_0^{\ast} +  \lambda_1^{\ast}\chi_t;
\lambda_{\xi}\left(\chi_{t},\xi_{t},\theta_{t},V_t\right)= \lambda_2^{\ast} +  \lambda_3^{\ast}\xi_t\\
&\lambda_{\theta}\left(\chi_{t},\xi_{t},\theta_{t},V_t\right)=
\lambda_5^{\ast} +  \lambda_4^{\ast}\sqrt{V_t};
\lambda_{V}\left(\chi_{t},\xi_{t},\theta_{t},V_t\right)= \lambda_6^{\ast} +  \lambda_7^{\ast}V_t\\
&\lambda_{0}=\lambda^{\ast}_{0}\sigma_{\chi};
\lambda_{1}=\lambda^{\ast}_{1}\sigma_{\chi};
\lambda_{2}=\lambda^{\ast}_{2}\sigma_{\xi};
\lambda_{3}=\lambda^{\ast}_{3}\sigma_{\xi}; \lambda_{4}=\lambda^{\ast}_{4}; \lambda_{5}=\lambda^{\ast}_{5}=0\\
&\lambda_{6}=\lambda^{\ast}_{6}\sigma_{V};
\lambda_{7}=\lambda^{\ast}_{7}\sigma_{V};
\beta^{\ast}=\beta+\lambda_1; \mu^{\ast}_{\xi}= \mu_{\xi} - \lambda_2\\
&\kappa^{\ast}_{\xi}= \kappa_{\xi}+\lambda_3;\mu^{\ast}_{V} =
\mu_{V} - \lambda_6; \kappa^{\ast}_{V} = \kappa_{V}+\lambda_7.
\end{align*}

Note the real processes are simply obtained by setting the risk premia ($\lambda$) parameters to zero. We can also ensure this model remains arbitrage free, according to discussions in \cite{cheridito2007market}, by considering constraints on the parameter $\mu_V$ such that $2 \mu_V \geq 1$ under both the real world and risk neutral measures, to ensure boundary non-attainment discussed in \cite{cheridito2007market}. This can be enforced in the prior specification we develop in Section 5. The parametric seasonal model we considered in this paper is given by:
\begin{equation}
f(t) = \sum^{12}_{k=2}\omega_{k}\Omega_{kt},
\label{seasonal}
\end{equation}
where $\Omega_{kt}=1$ if $t$ is the date of the $k$-th calendar month and $\Omega_{kt}=0$ otherwise. Note, without loss of generality one can set $f(0)=0$ as this constant is absorbed by $\xi_t$ and all variation is with respect to January.

\subsection{Commodity futures price.}
In this section we obtain analytic expressions for the futures price under the proposed multi-factor long-short dynamic model. The futures price we consider here will be developed as a function of the underlying stochastic factors. We denote the Futures contract price at time $t$ with maturity of the contract at $T$ by $F_{t,T}$ with the proof of this results provided in Appendix 1.\\
\textbf{Proposition 1:}\textit{ The futures contract price $F_{t,T}$ at time $t$ for a contract maturing at time $T$ has a closed from solution, with respect to the latent state variables that comprise the log spot price, $\chi_t, \xi_t, \theta_t$, when priced under the risk neutral measure, given by
\begin{equation*}
F_{t,T} = \exp\left(f(T) + B_0(\tau) + B_1(\tau)\xi_t +
B_2(\tau)\chi_t + B_3(\tau)\theta_t  \right),
\end{equation*}
with coefficients given by,
\begin{align*}
&B_0(\tau) = \frac{\sigma_{\xi}^2}{4\kappa_{\xi}^{\ast}}
\exp(-2\kappa_{\xi}^{\ast}\tau) +
\frac{\sigma_{\chi}^2}{4\beta^{\ast}}\exp(-2\beta^{\ast}\tau) +\\
&\frac{\mu_{\xi}^{\ast}}{\kappa_{\xi}^{\ast}}
\exp(-\kappa_{\xi}^{\ast}\tau) -
\frac{\lambda_0}{\beta^{\ast}}\exp(-\beta^{\ast}\tau) +
\frac{\rho_{\chi\xi}\sigma_{\chi}\sigma_{\xi}}{\beta^{\ast} +
\kappa_{\xi}^{\ast}}\exp(-\beta^{\ast}\tau -
\kappa_{\xi}^{\ast}\tau)
\\
&B_1(\tau) = \exp(-\kappa_{\xi}^{\ast}\tau); B_2(\tau)
=\exp(-\beta^{\ast}\tau); B_3(\tau) = 1.
\end{align*}
}\\
The proof of this result is provided in detail in Appendix 1. \\
\textbf{Remark 1:} \textit{ Note that the parameters $B_i(\tau)$ are under the risk neutral
framework and the latent factors in the futures price are real.}

Having derived the futures price for the stochastic model we propose, we now specify a state space model formulation and a Bayesian model. We will then develop a novel sampling methodology to estimate jointly the model parameters and
the latent factors.

\section{State Space Model via Strong Milstein Sheme: long-short stochastic volatility model.}
In specifying the state space models, we shall consider a panel of $N$ different futures contracts with maturities $T_1,\ldots,T_N$ and futures prices at time $t$ denoted by $F_{t,T_1},\ldots,F_{t,T_N}$. In addition, we note that the filtering framework is set up such that the latent process is the real process with real parameters used and the observation process is the real process with risk neutral parameters used.

\subsection{State Equations}
In this section we convert our s.d.e. models for the latent factors into a discretized version to obtain a state space model formulation. To achieve this we consider strong Taylor schemes, where an approximation process $Y$ converges in a strong sense with order $\gamma \in \left(0,\infty\right]$ with a continuous process $X$ if there exists a finite constant $K$ and a positive constant $\delta$ such that,
\begin{equation*}
\text{E}\left(|X_T-Y_N|\right) \preceq K\delta^{\gamma}.
\end{equation*}
We employ a strong stochastic Milstein Scheme. For the long-short factors the strong stochastic Taylor scheme of order 1, otherwise known as a Milstein scheme, corresponds to the Euler scheme.

For the volatility related stochastic processes the Milstein scheme is employed for these two processes since firstly they are both correlated, so the same discretization scheme should be employed for each of them. Secondly the process for $dV_t$ certainly does not have a Gaussian transition density, as such a Euler type scheme with additive Gaussian increments is too
simplistic an approximation, even at daily intervals. In the Milstein scheme we truncate the mixed integrals at a p-th order as shown in Appendix 2. 

\textbf{Proposition 2:} \textit{
The bivariate s.d.e.'s for $\chi_t$ and $\xi_t$ with 2-dimensional Wiener process discretized under a strong Euler scheme leads to the following state space equations.
\begin{align*}
&\chi_{t}= (1-\beta \triangle t)\chi_{t-1} +
\sigma_{\chi} \sqrt{\triangle t}n_{\chi}\\
&\xi_{t}= \mu_{\xi} + \left(1- \kappa_{\xi} \triangle t\right)\xi_{t-1} +
\sigma_{\xi} \sqrt{\triangle t} n_{\xi}
\end{align*}
where $n_{\chi}, n_{\xi}$ are standard normal random variables for the innovation noise.
The bivariate s.d.e.'s for $\theta_t$ and $V_t$ with 2-dimensional Wiener process discretized under a strong Milstein scheme leads to the following state space equations.
\begin{align*} 
&\theta_{t} \approx \theta_{t-1} + \sqrt{V_{t-1}\triangle t}n_{\theta,t} + \frac{1}{4}\sigma_V \rho_{V \theta} \left(\triangle t n^2_{\theta,t} - \triangle t\right) + \frac{1}{2}\sigma_V \sqrt{1-\rho^2_{V \theta}} J^p_{(2,1)\triangle t}\\
&V_t \approx V_{t-1} + \left(\mu_{V} - \kappa_{V}V_{t-1}\right) \triangle t + \sigma_V \rho_{V \theta} \sqrt{V_{t-1} \triangle t} n_{\theta,t} + \sigma_V \sqrt{V_{t-1} \left(1-\rho^2_{V \theta}\right) \triangle t} n_{V,t} \\
&+ \frac{1}{4}\sigma_V^2 \rho_{V \theta}^2 \left(\triangle t n^2_{\theta,t} - \triangle t \right) + 
\frac{1}{2}\sigma_V^2 \rho_{V \theta} \sqrt{1-\rho_{V \theta}^2} J^p_{(1,2) \triangle t} + 
\frac{1}{2}\sigma_V^2 \rho_{V \theta} \sqrt{1-\rho_{V \theta}^2} J^p_{(2,1) \triangle t} + 
\frac{1}{4}\sigma_V^2 \left(1-\rho_{V \theta}^2\right) \left(\triangle t n^2_{V,t} - \triangle t \right)
\end{align*}
where $n_{\theta,t}$ and $n_{V,t}$ are i.i.d. standard normal random variables and we utilise a $p$-th order truncation with terms for $J^p_{(i,j) \triangle t}$ defined by
\begin{equation*}
\begin{split}
J_{(j_1,j_2)\triangle t}^p &= \triangle t \left(
\frac{1}{2}\zeta_{j_1}\zeta_{j_2} +
\sqrt{\rho_{p}}\left(\mu_{j_1,p}\zeta_{j_2} -
\mu_{j_2,p}\zeta_{j_1}\right)\right) 
+ \frac{\triangle t}{2 \pi} \sum_{r=1}^p
\frac{1}{r}\left(\psi_{j_1,r}\left(\sqrt{2}\zeta_{j_2} +
\nu_{j_2,r}\right) - \psi_{j_2,r}\left(\sqrt{2}\zeta_{j_1} +
\nu_{j_1,r}\right)\right),
\end{split}
\label{eqn:Jmix}
\end{equation*}
where $\zeta_{j},\mu_{j,p},\nu_{j,r}$ and $\psi_{j,r}$ are all
independent $\mathcal{N}(0;1)$ Gaussian random variables with,
\begin{equation*}
\begin{split}
\rho_p &= \frac{1}{12} - \frac{1}{2 \pi^2}\sum_{r=1}^p\frac{1}{r^2}, \; \;
\zeta_j = \frac{1}{\sqrt{\triangle t}} \triangle W^j.
\end{split}
\end{equation*}
In the specification of our models we take $\triangle t = 1$ day. 
}\\
The proof of these results is presented in Appendix 2.

\textbf{Remark 2:} \textit{Under this discretization scheme it will not be possible in general to find the transition density for these two factors $\theta_t$ and $V_t$, given $\theta_{t-1}$ and $V_{t-1}$. We note that the version of the particle filter we develop only requires that we can generate from these state equations, which is trivial in the sense that it only requires generating Gaussian random variables and applying a transform. This is the key reason for our use later on of the SIR particle filter approach in our estimation methodology.}

\subsection{Observation Equations}
\label{observations}
Clearly, the observation equations for the futures contracts are
linear and will be given in Equations \ref{FuturesObsEqns}.
\begin{align*}
\ln F_{t,T_1} &= f(T_1) + B_0(\tau_1) + B_1(\tau_1)\xi_t +
B_2(\tau_1)\chi_t + B_3(\tau_1)\theta_t + \epsilon_t^{(1)} \\
\vdots\\
\ln F_{t,T_N} &= f(T_N) + B_0(\tau_N) + B_1(\tau_N)\xi_t +
B_2(\tau_N)\chi_t + B_3(\tau_N)\theta_t + \epsilon_t^{(N)}
\label{FuturesObsEqns}
\end{align*}
where $\epsilon_{t}^{(1)}, \ldots, \epsilon_{t}^{(N)}$ are all zero mean normal
random variables for the observation noise on the futures contract prices, with covariance structures at each time $t$ denoted by $\Sigma_{\epsilon}$ which is $(N \times N)$ dimensional.

Clearly in this model the state equations, presented in Proposition 2, are non-linear and non-Gaussian, hence the use of a Kalman-Filter is no-longer optimal. It is also clear that a simple linearization of the state equations will not suffice, hence this also precludes the use of the approximate Extended Kalman Filter formulation. Instead we will develop a particle filter and Markov chain Monte Carlo solution based around a novel version of the Particle MCMC algorithm.

\section{Bayesian Model for Commodities}
\label{BayesModel}
In this section we develop a Bayesian model for the parameters and latent factors that must be estimated over time. In introducing a Bayesian formulation for this problem we note that we model the real process parameters as random variables and the risk premia used to convert to the risk neutral pricing framework as unknown random variables. The Bayesian formulation proceeds by first defining the posterior distribution one is interested in considering:
\begin{align}
\pi&\left(\bm{\Phi},\chi_{1:t},\xi_{1:t},\theta_{1:t},V_{1:t}|F_{1:t,T_1},\ldots,F_{1:t,T_N}\right)\\ 
&=\mu_{\bm{\Phi}}(\chi_{0},\xi_{0},\theta_{0},V_{0})\prod_{t=1}^{T}f^{(\chi)}_{\bm{\Phi}}\left(\chi_{t}|\chi_{t-1}\right)f^{(\xi)}_{\bm{\Phi}}\left(\xi_{t}|\xi_{t-1}\right)
f^{(\theta)}_{\bm{\Phi}}\left(\theta_{t}|\theta_{t-1},V_{t-1}\right)f^{(V)}_{\bm{\Phi}}\left(V_{t}|V_{t-1}\right)\\
\times
&\prod_{t=1}^{T}\prod_{n=1}^{N}g_{\bm{\Phi}}(F_{t,T_n}|\chi_{t},\xi_{t},\theta_{t}) p(\bm{\Phi})
\label{eqn:Post}
\end{align}
where we define quantities as follows:
\begin{itemize}
\item{$\bm{\Phi} = \left(\bm{\Phi_R},\bm{\Phi_{RN}},\omega_{1:11}\right)$, where $\bm{\Phi_R}$ and $\bm{\Phi_{RN}}$ are the real and risk neutral parameter vectors respectively, e.g. $\bm{\Phi_R} = \left(\beta,\mu_{\xi},\kappa_{\xi},\mu_V,\kappa_V,\sigma_{\chi},\sigma_{\xi},\sigma_{V},\Sigma_{\epsilon}\right)$ the random vector of unknown real-world model parameters.}
\item{\textit{A priori} there is no reason to specify the priors for the real-world and risk-neutral parameters differently and therefore we give them identical prior specifications. With the real-world prior $p(\bm{\Phi_R})$ for the random parameter vector $\bm{\Phi_R}$ given by:
\begin{align*}
&\beta_{\chi} \sim U\left(A_{\beta_{\chi}},B_{\beta_{\chi}}\right); \sigma^2_{\chi} \sim U(A_{\sigma^2_{\chi}},B_{\sigma^2_{\chi}}); \mu_{\zeta} \sim U\left(M_{\mu_{\zeta}},V_{\mu_{\zeta}}\right); \kappa_{\zeta} \sim U\left(A_{\kappa_{\zeta}},B_{\kappa_{\zeta}}\right); \sigma^2_{\zeta} \sim U(A_{\sigma^2_{\zeta}},B_{\sigma^2_{\zeta}}); \\
&\mu_{V} \sim U\left(A_{\mu_{V}},B_{\mu_{V}}\right); \kappa_{V} \sim U\left(A_{\kappa_{V}},B_{\kappa_{V}}\right); \sigma^2_{V} \sim U(A_{\sigma^2_{V}},B_{\sigma^2_{V}});\sigma^2_{F(T_i)} \sim IG(A_{\sigma^2_{F(T_i)}},B_{\sigma^2_{F(T_i)}}).
\end{align*}
With respect to the choices of hyper-prior parameters, we consider an objective uninformative prior specification, see Section 7. Furthermore, we assume that $\Sigma_{\epsilon} = \text{diag}\left(\sigma^2_{\epsilon_1},\ldots,\sigma^2_{\epsilon_N}\right)$, though it is trivial to consider alternative specifications.}
\item{$\mu_{\bm{\Phi}}(\chi_{0},\xi_{0},\theta_{0},V_{0})$ is a multivariate Gaussian with known mean $\bm{\nu}$ and covariance $\Sigma_{\nu}$ which specifies the distribution of the initial state of the factors given the model parameters.}
\item{$f^{(\chi)}_{\bm{\Phi}}\left(\chi_{t}|\chi_{t-1}\right)$ is given by the state space model state
equation to be a Gaussian with mean $(1-\beta)\chi_{t-1}\triangle
t$ and standard deviation $\sigma_{\chi}\sqrt{\triangle t}$.}
\item{$f^{(\xi)}_{\bm{\Phi}}\left(\xi_{t}|\xi_{t-1}\right)$ is given by the state space model state
equation to be a Gaussian with mean $\mu_{\xi}\triangle t +
(1-\kappa_{\xi}\triangle t)\chi_{t-1}$ and standard deviation
$\sigma_{\xi}\sqrt{\triangle t}$.}
\item{$f^{(\theta)}_{\bm{\Phi}}\left(\theta_{t}|\theta_{t-1}\right)$ and $f^{(V)}_{\bm{\Phi}}\left(V_{t}|V_{t-1}\right)$ are given by the state space model state equations to be the distribution formed from the addition of the
random variables $\sqrt{V_{t-1}}n_{\theta,t} + e_{\theta,t}^{1,1}
+ e_{\theta,t}^{2,1}$ and $\sigma_{V}\sqrt{V_{t-1}} n_{V,t} +
h_{V,t}^{2,2}$ respectively. The convolution integrals involved with
finding these distributions do not admit a parametric form. We
 that this does not affect our estimation approach since we only
require that we can sample efficiently these processes.}
\item{$g_{\bm{\Phi}}(F_{t,T_n}|\chi_{t},\xi_{t},\theta_{t})$ is given by the state space model observation equation to be a Gaussian with mean $f(T_1) + B_0(\tau) +
B_1(\tau)\xi_t + B_2(\tau)\chi_t + B_3(\tau)\theta_t$ and
standard deviation $1$.}
\end{itemize}

\section{Parameter estimation and filtering via Adaptive MCMC within Rao-Blackwellised Particle MCMC}
In this section we detail the simulation methodology we develop for obtaining samples from the posterior distribution given in Equation (\ref{eqn:Post}). This simulation methodology is based on a recent sampler specifically developed for use in state space models in which one wishes to estimate jointly the model parameters and the latent states, see \cite{AndrieuDoucetHolenstein2010}. The methodology is known as Particle Markov chain Monte Carlo (PMCMC) and represents a state of the art sampling framework for such problems. The approach embeds a particle filter estimate of an optimal proposal distribution into a MCMC algorithm, allowing one to obtain samples from the target posterior distribution for the
parameters and latent states. This embedding is done in a non-trivial fashion, see \cite{AndrieuDoucetHolenstein2010} for details.

The Particle MCMC proposal distribution is split into two components. The first involves a proposal kernel which is constructed via an adaptive Metropolis scheme and this will be used to sample the static parameters $\bm{\Phi}$. The second component of the proposal kernel involves the sampling of a trajectory for the latent factors, $\chi_{1:t},\xi_{1:t},\theta_{1:t},V_{1:t}$. This proposal kernel is constructed via a Rao-Blackwellized Sequential Monte Carlo algorithm in which the Rao-Blackwellizing variance reduction is achieved via optimal Kalman-Filtering for the path space associated to $\chi_{1:t},\xi_{1:t}$.

The introduction of an adaptive MCMC proposal kernel into the Particle MCMC setting allows the Markov chain proposal distribution to adaptively learn the regions in which the marginal posterior distribution for the static model parameters has most mass. As such the probability of acceptance under such an online adaptive proposal will be significantly improved over time. Then for the latent factor path spaces we develop a minimum variance approach involving Rao-Blackwellisation via a Kalman Filter, followed by a Sampling Importance Resampling (SIR) version of the particle filter. \textit{Note - it is important to work with an SIR filter for the path space parameters associated with $\theta_{1:t},V_{1:t}$. The reason for this is that under the Milstein discretization scheme we develop, we can efficiently sample exactly from the prior parameters but can not evaluate the prior density point-wise.} Hence, under this variance reduction sampling framework, we optimally sample the latent state path space trajectories for $\chi_{1:t}$ and $\xi_{1:t}$ via a Kalman Filter formulation and we use a SIR version of the particle filter for the state path trajectories for $\theta_{1:t},V_{1:t}$.

\subsection{Particle Markov chain Monte Carlo (PMCMC) Background}
Particle MCMC is a new methodology introduced recently in \cite{AndrieuDoucetHolenstein2010}. The methodology utilizes particle filtering methodology to approximate the optimal proposal distribution in an MCMC algorithm in which one is interested in high-dimensional block updates. This naturally, lends it self well to the setting of state space modelling. In the setting considered in this paper the parameter space of static parameters in the model is high dimensional, and in addition the path space latent factors given an additional $4t$ parameters. Where, in this case $t$ can range from hundreds to thousands of days, depending on the commodities studied and sampling frequency of the observations. Hence, this problem is ideally suited for the PMCMC methodology.

The simplest solution to this problem would be to sample directly form the univariate full conditional distributions. However, even in the case in which a single component Gibbs sampler is achievable via inversion sampling for all parameters, this sampling scheme will typically results in very slow mixing of the Markov chain around the support of the posterior. This is especially problematic in high dimensional target posterior distributions, since it requires excessively long Markov chain runs to achieve samples from stationary regime. Typically, it can also lead to very high autocorrelations in the Markov chain states.

It is well known that to avoid this slow mixing Markov chain setting, one must sample from larger blocks of parameters. However, the design of an optimal proposal distribution for large blocks of parameters is very complicated. The Particle MCMC methodology allows one to approximate the optimal proposal distribution for a large number of parameters via a Sequential Monte Carlo approximated proposal distribution.

In the state space setting, the Particle MCMC algorithm used to sample from a generic target distribution, with static parameters $\bm{\theta} = \left(\bm{\Phi_R},\bm{\Phi_RN}\right)$ and latent state vector $\bm{x}_{t}=\left(\chi_t,\xi_t,V_t,\theta_t\right)$, $\pi\left(\bm{\theta},\bm{x}_{1:t}|\bm{y}_{1:t}\right)$ proceeds by mimicking the marginal Metropolis-Hastings algorithm in which the acceptance probability is given by,
\begin{align*}
\alpha\left(\bm{\theta},\bm{\theta}'\right) = \min \left( 1,
\frac{\pi\left(\bm{\theta}'|\bm{y}_{1:t}\right)q\left(\bm{\theta}',\bm{\theta}\right)}{\pi\left(\bm{\theta}|\bm{y}_{1:t}\right)q\left(\bm{\theta},\bm{\theta}'\right)}
\right).
\end{align*}
Clearly, achieving this requires one to use a very particular structure for the proposal kernel in the MCMC algorithm. In particular the proposal kernel must take the form,
\begin{align*}
q\left(\left(\bm{\theta},\bm{x}_{1:t}\right),\left(\bm{\theta}',\bm{x}'_{1:t}\right)\right)
=q\left(\bm{\theta}',\bm{\theta}\right)\pi\left(\bm{x}'_{1:t}|\bm{y}_{1:t},\bm{\theta}'\right).
\end{align*}
Under this proposal, the acceptance probability for a standard Metropolis-Hastings algorithm takes the form,
\begin{align*}
\alpha\left(\bm{\theta},\bm{\theta}'\right) &= \min \left( 1,
\frac{\pi\left(\bm{\theta}',\bm{x}'_{1:t}|\bm{y}_{1:t}\right)q\left(\left(\bm{\theta},\bm{x}_{1:t}\right),\left(\bm{\theta}',\bm{x}'_{1:t}\right)\right)}
{\pi\left(\bm{\theta},\bm{x}_{1:t}|\bm{y}_{1:t}\right)q\left(\left(\bm{\theta}',\bm{x}'_{1:t}\right),\left(\bm{\theta},\bm{x}_{1:t}\right)\right)}
\right)\\
&= \min \left( 1,
\frac{\pi\left(\bm{\theta}'|\bm{y}_{1:t}\right)q\left(\bm{\theta}',\bm{\theta}\right)}{\pi\left(\bm{\theta}|\bm{y}_{1:t}\right)q\left(\bm{\theta},\bm{\theta}'\right)}
\right).
\end{align*}

The critical idea recognized in \cite{AndrieuDoucetHolenstein2010} in formulating the
Particle MCMC algorithm is that the proposal distribution for $\pi\left(\bm{x}'_{1:t}|\bm{y}_{1:t},\bm{\theta}'\right)$ can be sampled from via a Sequential Monte Carlo algorithm.

\subsubsection{Generic Particle MCMC Algorithm}
One iteration of the generic Particle MCMC algorithm proceeds as follows:
\begin{enumerate}
\item{Sample $\bm{\theta}' \sim q\left(\bm{\theta},\cdot\right)$.}
\item{Run an SMC algorithm with $N$ particles to obtain:
\begin{align*}
&\widehat{\pi}\left(\bm{x}_{1:t}|\bm{y}_{1:t},\bm{\theta}'\right)
=\sum_{i=1}^{N}W_{1:t}^{(i)}\delta_{\bm{x}_{1:t}^{(i)}}\left(x_{1:t}\right)\\
&\widehat{\pi}\left(\bm{y}_{1:t}|\bm{\theta}'\right) =
\prod_{j=1}^t \left(\frac{1}{N}\sum_{i=1}^{N}
w_j\left(\bm{x}_{j}^{(i)}\right) \right)
\end{align*}
Then sample a candidate path $\bm{X}'_{1:t} \sim
\widehat{\pi}\left(\bm{x}_{1:t}|\bm{y}_{1:t},\bm{\theta}'\right)$.}
\item{Utilise the estimate of $\widehat{\pi}\left(\bm{y}_{1:t}|\bm{\theta}\right)$ from the previous iteration of the Markov chain.}
\item{Accept the proposed new Markov chain state comprised of
$\left(\bm{\theta}',\bm{X}'_{1:t}\right)$ with acceptance probability given by
\begin{align}
&\alpha\left(\left(\bm{\theta},\bm{X}_{1:t}\right),\left(\bm{\theta}',\bm{X}'_{1:t}\right)\right)
 = \min \left( 1, \frac{\widehat{\pi}\left(\bm{y}_{1:t}|\bm{\theta}'\right)\pi\left(\bm{\theta}'\right)q\left(\bm{\theta}',\bm{\theta}\right)}
 {\widehat{\pi}\left(\bm{y}_{1:t}|\bm{\theta}\right)\pi\left(\bm{\theta}\right)q\left(\bm{\theta},\bm{\theta}'\right)} \right)
\end{align}
}
\end{enumerate}

The key advantage of this approach is that the difficult problem of designing high dimensional proposals has been replaced with the simpler problem of designing low dimensional mutation kernels in
the Sequential Monte Carlo algorithms embedded in the MCMC algorithm. In the paper \cite{AndrieuDoucetHolenstein2010}, this sampling approach in which Sequential Monte Carlo is used to approximate the marginal likelihood in the acceptance probability has been shown to have several theoretical convergence properties, see Section 4 of \cite{AndrieuDoucetHolenstein2010} for details. 

\subsection{Adaptive Markov chain Monte Carlo (AdMCMC) Background}
There are several classes of adaptive MCMC algorithms, see \cite{Roberts08}. The distinguishing feature of adaptive MCMC algorithms, compared to standard MCMC, is generation of the Markov chain via a sequence of transition kernels. Adaptive algorithms utilise a combination of time or state inhomogeneous proposal kernels. Each proposal in the sequence is allowed to depend on the past history of the Markov chain generated, resulting in many possible variants.

Due to the inhomogeneity of the Markov kernel used in adaptive algorithms, it is particularly important to ensure the generated Markov chain is ergodic, with the appropriate stationary
distribution. Several recent papers proposing theoretical conditions that must be satisfied to ensure ergodicity of adaptive algorithms include,\cite{Atachade05},\cite{Roberts08}, \cite{Haario07}, \cite{Andrieu06} and \cite{Andrieu07}.

In \cite{Haario01} an adaptive Metropolis algorithm with proposal covariance adapted to the history of the Markov chain was developed. The original proof of ergodicity of the Markov chain under such an adaption was overly restrictive. It required a bounded state space and a uniformly ergodic Markov chain. In \cite{Roberts08} a proof of ergodicity of adaptive MCMC under simpler conditions known as \textit{Diminishing Adaptation} and \textit{Bounded Convergence} is presented. In general it is non-trivial to develop adaption schemes which can be verified to satisfy these two conditions. In this paper we use the adaptive MCMC algorithm to learn the proposal distribution for the static parameters in our posterior $\bm{\Phi}$. In particular we work with an Adaptive Metropolis algorithm utilizing a mixture proposal kernel known to satisfy these two ergodicity conditions for unbounded state spaces and general classes of target posterior distribution, see \cite{Roberts08} for details.

\subsubsection{Adaptive Metropolis within Rao-Blackwellised Particle MCMC}
In this section we present the specific details of the Adaptive Metropolis within Rao-Blackwellised Particle MCMC algorithm used to sample from the posterior on the path space of our latent factors and state space model parameters. This involves specifying the details of the proposal distribution in the PMCMC algorithm,
{\small{
\begin{align*}
q\left(\left(\bm{\Phi},\chi_{1:t},\xi_{1:t},\theta_{1:t},V_{1:t}\right),\left(\bm{\theta}',\chi'_{1:t},\xi'_{1:t},\theta'_{1:t},V'_{1:t}\right)\right)
=q\left(\bm{\Phi}',\bm{\Phi}\right)\pi\left(\chi'_{1:t},\xi'_{1:t},\theta'_{1:t},V'_{1:t}|\bm{F_{1:t,T}},\bm{C_{1:t,T,M,K}},\bm{\Phi}'\right).
\end{align*}
}}
The proposal, $q\left(\bm{\Phi}^{(j-1)},\bm{\Phi}'\right)$, for the static model parameters $\bm{\Phi}$ is given by an adaptive Metropolis proposal comprised of a mixture of Gaussians, one component of which has a covariance structure which is adaptively learnt on-line as the algorithm progressively explores the posterior distribution. The mixture proposal distribution for parameters $\bm{\Phi}$ is given at iteration $j$ of the Markov chain by,
\begin{equation}
q_j\left(\bm{\Phi}^{(j-1)},\cdot\right)= w_1
N\left(\bm{\Phi};\bm{\Phi}^{(j-1)},\frac{\left(2.38\right)^2}{d}\Sigma_j\right)
+ \left(1-w_1\right)
N\left(\bm{\Phi};\bm{\Phi}^{(j-1)},\frac{\left(0.1\right)^2}{d}I_{d,d}\right).
\end{equation}
Here, $\Sigma_j$ is the current empirical estimate of the covariance between the parameters of $\bm{\Phi}$ estimated using samples from the Particle Markov chain up to time $j$. The theoretical motivation for the choices of scale factors 2.38, 0.1 and dimension $d$ are all provided in \cite{Roberts08} and are based on optimality conditions presented in \cite{Roberts97} and \cite{Roberts01}. We note that the update of the covariance matrix, can be done recursively online via the following recursion,
\begin{equation}
\begin{split}
\mu_{j+1} &= \mu_{j} + \frac{1}{j+1}\left(\tilde{\beta}^{(j-1)} - \mu_j\right)\\
\Sigma_{j+1} &= \Sigma_{j} + \frac{1}{j+1}\left(\left(\tilde{\beta}^{(j-1)} - \mu_j\right)\left(\tilde{\beta}^{(j-1)} - \mu_j\right)' - \Sigma_j\right).
\end{split}
\end{equation}
Having specified the adaptive PMCMC proposal distribution for the static model parameters, we now focus on the proposal mechanism for the latent path space trajectories for the factors in the commodity model in the next subsection.

\subsection{Rao-Blackwellised Sequential Importance Sampling proposal}
\label{RaoSIRProp}
The proposal kernel for the latent factors that will give us a Marginal Metropolis-Hastings framework is given by, $\pi\left(\chi'_{1:t},\xi'_{1:t},\theta'_{1:t},V'_{1:t}|\bm{F_{1:t,T}},\bm{\Phi}'\right)$. Sampling from this proposal can not be done exactly as it corresponds to sampling from the full conditional posterior distribution for the latent factors path genealogies. Instead we will utilise a specifically designed Sequential Monte Carlo algorithm to construct our proposal distribution. 

The SMC algorithm we develop here has a particular form which allows us to achieve two goals, the first that we can perform Rao-Blackwellisation via a Kalman-Filter for the latent factors $\chi_{1:t},\xi_{1:t}$. This optimal marginalised particle filtering approach and ideas are presented in \cite{Doucet00} and similar ideas are discussed in \cite{Casella96}. Secondly, we develop a filter known as the Sequential Importance Sampling Resampling (SIR) filter for the latent factors $\theta_{1:t}, V_{1:t}$. In doing this we can sample exactly from the prior for our latent factors (i.e. the Milstein discretization) and perform the importance weighting only dependent on the likelihood model. The algorithm for the Adaptive Metropolis within Rao-Blackwellised Particle MCMC is presented in Appendix \ref{IRAlgs}. Where the details of step 5. are also provided in Appendix \ref{IRAlgs}. The details of step 6. pertain to the Rao-Blackwellised SIR particle filter for constructing the proposal for the path spaces of the latent factors in the model. This requires some additional discussion and is presented below in Section \ref{RaoSIRProp}.

\textbf{Remark 3:}\textit{ Development of a Rao-Blackwellized SIR particle filter proposal for the latent path trajectories proposal in the PMCMC algorithm, given the proposal static parameters $\bm{\Phi}$ has the advantage that it significantly reduces the variance in importance sampling weights. Achieved by exploiting the conditional linear Gaussian stat-space structure and a Kalman Filter. This will in turn increase the acceptance rate of our proposal mechanism.}

\textbf{Remark 4:}\textit{ Utilising the SIR filter to construct a proposal for the latent factors $\theta_{1:t}, V_{1:t}$ has an important advantage in the Milstein discretized state space model formulation we developed in that it can sample exactly from the prior for our latent factors and perform the importance weighting only dependent on the likelihood model. This is clearly beneficial in our setting since it allows us to avoid having to perform point-wise evaluation of the transition distributions given by $f_{\bm{\Phi}}\left(\theta_t|\theta_{t-1}\right)$ and $f_{\bm{\Phi}}\left(V_t|V_{t-1}\right)$ which can only be expressed as convolution integrals.} 

The specific algorithm used to complete Step 7 of the Adaptive Metropolis within Rao-Blackwellised Particle MCMC algorithm is provided in Section \ref{IRAlgs}. Here we present the details for the sampling of a candidate proposal, conditional on the proposed state $\bm{\Phi}'$, for the latent factor path spaces, $\chi_{1:t},\xi_{1:t},\theta_{1:t},V_{1:t}$ as well as the estimation of the marginal likelihood $\widehat{p}\left(\bm{F_{1:t,T}}|\bm{\Phi}'\right)$ required for the calculation of the acceptance probability. This involves the development of a Rao-Blackwellised Kalman Filter scheme in conjunction with an SIR particle filter stage. 

The Rao-Blackwellisation via the Kalman Filter involves the following decomposition of the filtering distribution,
\begin{align*}
\pi\left(\chi_{1:t},\xi_{1:t},\theta_{1:t},V_{1:t}|\bm{F_{1:t,T}},\bm{\Phi}\right)
&=\pi\left(\chi_{1:t},\xi_{1:t}|\theta_{1:t},V_{1:t},\bm{F_{1:t,T}},\bm{\Phi}\right)\\
&\times
\pi\left(\theta_{1:t},V_{1:t}|\bm{F_{1:t,T}},\bm{\Phi}\right)
\end{align*}
Next we note that the posterior distribution
$\pi\left(\chi_{1:t},\xi_{1:t}|\theta_{1:t},V_{1:t},\bm{F_{1:t,T}},\bm{\Phi}\right)$
is exactly characterized by the Kalman Filtering estimate. We seek
to exploit this fact in developing our sampling methodology. First we utilize a
standard SIR particle filter with adaptive resampling to obtain the particle estimate,
\begin{align*}
\widehat{\pi}\left(\theta_{1:t},V_{1:t}|\bm{F_{1:t,T}},\bm{\Phi}\right)
=\sum_{i=1}^{N}W_t^{(i)}\delta_{\left(\theta^{(i)}_{1:t},V^{(i)}_{1:t}\right)}\left(\theta_{1:t},V_{1:t}\right)
\end{align*}
This particle estimate of the marginal then results in the joint posterior distribution estimate,
\begin{align*}
\widehat{\pi}\left(\chi_{1:t},\xi_{1:t},\theta_{1:t},V_{1:t}|\bm{F_{1:t,T}},\bm{\Phi}\right)
=\sum_{i=1}^{N} W_t^{(i)}
\pi\left(\chi_{1:t},\xi_{1:t}|\theta^{(i)}_{1:t},V^{(i)}_{1:t},\bm{F_{1:t,T}},\bm{\Phi}\right).
\end{align*}
Now, clearly we can obtain
$\pi\left(\chi_{1:t},\xi_{1:t}|\theta^{(i)}_{1:t},V^{(i)}_{1:t},\bm{F_{1:t,T}},\bm{\Phi}\right)$,
for each particle $i$, an estimate of this distribution, optimally via the Kalman Filter. In addition, it is easily shown that in this version of the SIR particle filter the unnormalised importance sampling weight for particle $i$ is given by, see \cite{Doucet00},
\begin{align*}
\tilde{W}^{(i)}_t \propto
W^{(i)}_{t-1}p\left(\bm{F_{t,T}}|\bm{F_{1:t-1,T}},\bm{\Phi},\theta^{(i)}_{1:t},\chi^{(i)}_{1:t},\xi^{(i)}_{1:t}\right).
\end{align*}
We explain now the details of the Rao-Blackwellisation within the context of the SIR particle filter and then present the algorithm to perform this methodology. Given we have obtained a particle estimate $\{\theta^{(i)}_{1:t},V^{(i)}_{1:t},W_t^{(i)}\}_{i=1:N}$
at time $t$. We associate to each particle a Kalman Filter mean and covariance $\bm{\mu}_t^{(i)} =
\left(\mu^{(i)}_{\chi}(t),\mu^{(i)}_{\xi}(t)\right)$ and
$\bm{\Sigma}_{t}^{(i)} = \Sigma^{(i)}_{\chi,\xi}(t)$, which we obtain via a standard Kalman Filter recursion as follows:
\begin{align*}
&\bm{\mu}^{(i)}_{t|t-1} = \bm{A}_n\bm{\mu}^{(i)}_{t-1|t-1} \\
&\Sigma^{(i)}_{t|t-1} =
\bm{A}_n\Sigma^{(i)}_{t-1|t-1}\bm{A}^{\top}_n +
\sigma^2\bm{b}_n\bm{b}^{\top}_n \\
&\Gamma_t^{(i)} = \bm{b}_n^{\top}\Sigma^{(i)}_{t|t-1}\bm{b}_n\\
&y^{(i)}_{t|t-1} = \bm{b}_n^{\top}\bm{\mu}^{(i)}_{t|t-1} +
\bm{b}^{\top}_s\left(\theta_t^{(i)},V_t^{(i)}\right) \\
&\bm{\mu}^{(i)}_{t|t} = \bm{\mu}^{(i)}_{t|t-1} -
\Sigma^{(i)}_{t|t-1}\bm{b}_n\left[\Gamma_t^{(i)}\right]^{-1}\left(y_t
- y^{(i)}_{t|t-1}\right)\\
&\Sigma^{(i)}_{t|t} = \Sigma^{(i)}_{t|t-1} -
\Sigma^{(i)}_{t|t-1}\bm{b}_n\left[\Gamma_t^{(i)}\right]^{-1}\bm{b}_n^{\top}\Sigma^{(i)}_{t|t-1},
\end{align*}
where $\Gamma_t =
\text{Cov}\left(\bm{F_{t,T}}|\bm{F_{1:t-1,T}}\right)$
and the predictive density is given by
\begin{align*}
p\left(\bm{F_{t,T}}|\bm{F_{1:t-1,T}},\bm{\Phi},\theta_{1:t},\chi_{1:t},\xi_{1:t}\right)=\mathcal{N}\left(\bm{F_{t,T}};f_1\left(\bm{F_{1:t-1,T}}\bm{\Phi},\theta_{1:t},\chi_{1:t},\xi_{1:t}\right),\Gamma^{(1)}_t\right).
\end{align*}

Now that we have detailed each aspect of the model and each aspect of the proposed sampling algorithm to allow us to calibrate and estimate the Bayesian commodity model developed, we will now present the results and analysis. This will be done in two stages, first we present details of a comprehensive synthetic simulation study, followed by a real data analysis.

\section{Results and Discussion}
In this section we present studies on the performance of the Bayesian model formulation, the Milstein discretization framework and the Adaptive PMCMC sampling framework under Rao-Blackwellization. To achieve this we first consider a case study consisting of synthetic data analysis. After demonstrating the accuracy of the estimation and calibration methodology we developed, we perform a detailed analysis utilising actual data representing a panel of crude oil futures.

\subsection{Synthetic Data Example}
In this section we present detailed analysis of the performance of the Adaptive PMCMC algorithm in the context of the Bayesian state space model non-linear filtering and parameter estimation framework that we have developed. This study aims to systematically assess the following model and AdPMCMC sampler aspects: the ability to estimate static model parameters (perform calibration) in high observation noise; the impact of parameter estimation under vague priors for static parameters corresponding to drift and volatility of each of the four discretized s.d.e. factor models; the impact of increasing the number of futures contracts with different maturities on parameter and latent factor estimation; and finally, the impact of the Milstein scheme truncation for $p$.

\subsubsection{Analysis of Estimation Accuracy versus Number of Futures Contracts}
\begin{figure}[!ht]
\includegraphics[width = \textwidth, height = 10cm]{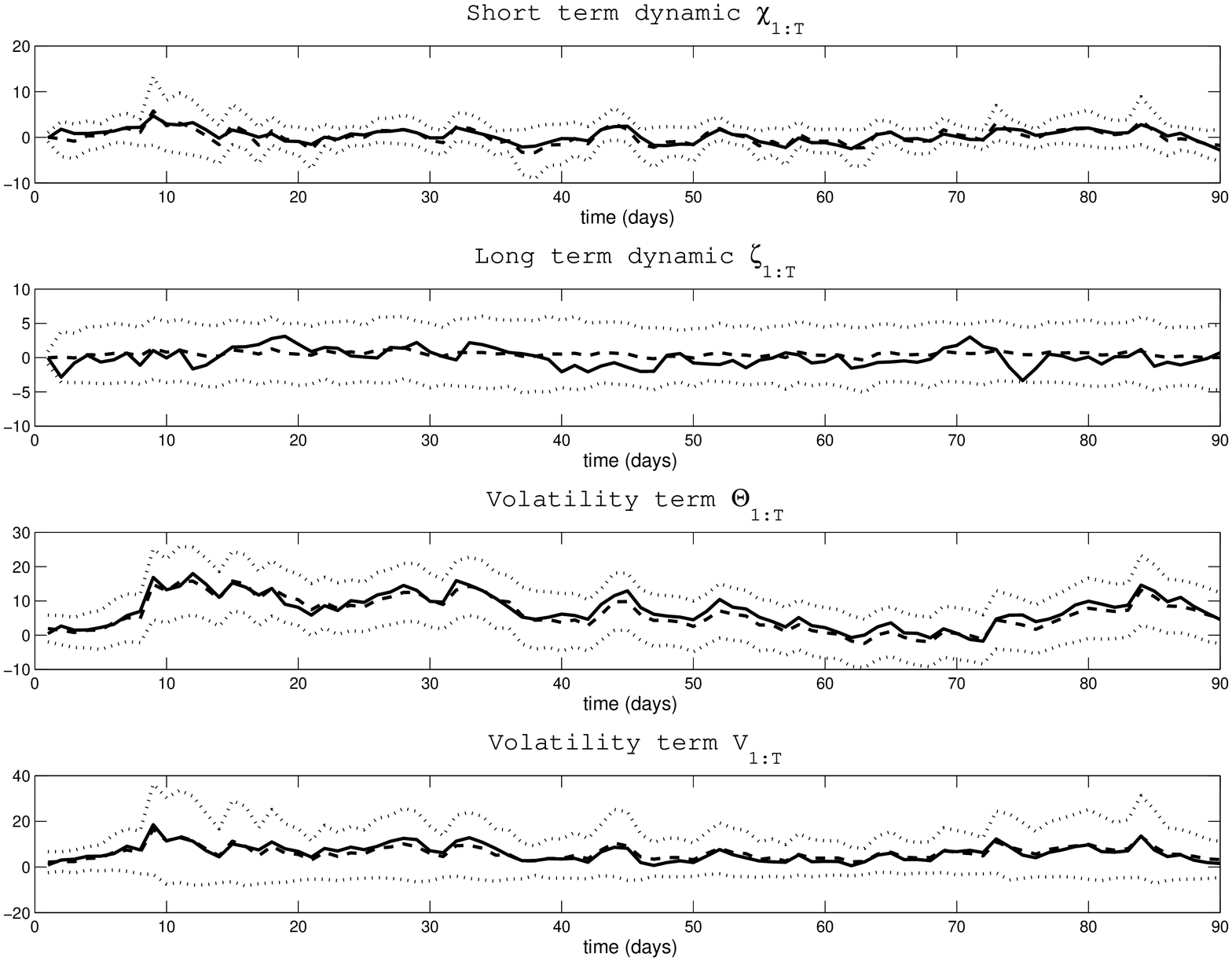}
\caption{True latent factor model s.d.e. trajectories (solid line) versus Rao-Blackwellised AdPMCMC sampler filter estimates for the MMSE (dashed line) and 95\% posterior confidence intervals (dotted lines) for each factor: short term and long term trends and volatility.  The Milstein SIR proposal utilised $p=100$ and 500 particles.}
\label{FigDynamicFactors}
\end{figure}

\begin{figure}[!ht]
\includegraphics[width = \textwidth, height = 8cm]{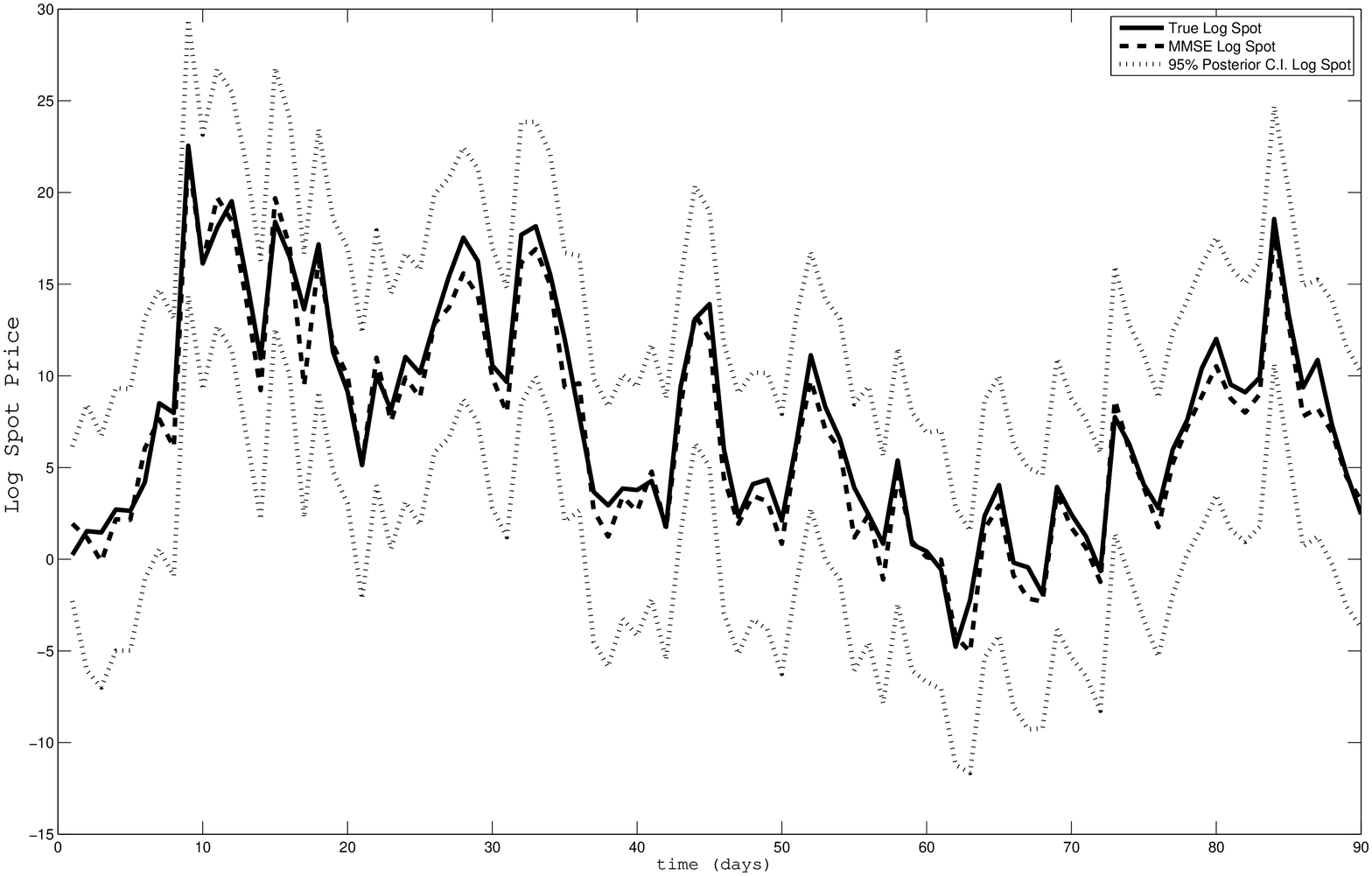}
\caption{True latent log spot price trajectory versus Rao-Blackwellised AdPMCMC sampler filter estimated log spot price MMSE and 95\% posterior confidence interval.}
\label{FigDynamicSpot}
\end{figure}

The simulation study for this section proceeded by first sampling a set of static model parameters from the priors which we will utilise to generate synthetic data. We will then obtain posterior estimates of these parameters given the generated data. The `true' static parameters utilised in this study, were given by $\bm{\Phi_R} = \left(\beta,\mu_{\xi},\kappa_{\xi},\mu_V,\kappa_V,\sigma_{\chi},\sigma_{\xi},\sigma_{V},\Sigma_F\right) = \bm{\Phi_{RN}} = \left(0.2,0.1,0.4,0.2,0.2,0.5,0.5,0.5,4\right)$ and seasonality components $w_{1:11} = 0$. These parameter settings were then utilized to generate a set of latent process trajectories which were then in turn utilized to generate a set of futures curves with observations in noise. This produced a set of synthetic latent paths, for 100 days $T=100$, for $\chi_{1:T}, \zeta_{1:T}, \theta_{1:T}, V_{1:T}$ which are presented in the upper panel of Figure \ref{FigDynamicFactors}. The lower panel of Figure \ref{FigDynamicSpot} presents the resulting log spot price each day. We then utilised these simulated trajectories to generate a panel of futures curves according to the observation equations developed in Section \ref{observations}. We utilised a panel of futures curves corresponding to 10 futures contracts, each separated by maturities of 30 days, the first contract rolling over in 29 days time, followed by the next contract maturing in 59 days, all the way up to the tenth contract which at time $t=1$ of the observations still had 299 days remaining until maturity. Using these latent process, the noisy observations of the futures curve was generated over 100 days of observations using $\Sigma_F = \text{diag}\left(\sigma^2_{F(T_1)},\ldots,\sigma^2_{F(T_N)}\right) = \left(1,\ldots,1\right)$. Additionally, unless otherwise stated we ran the AdPMCMC sampler presented in Algorithm 1 for $J = 200,000$ Markov chain samples with $N = 500$ particles. The discard Markov chain samples for burn-in was selected as $50,000$.  The discretization interval considered was daily, $\triangle t = 1$ and the p-th order truncation of the Milstein discretization used to calculate $\rho_p$ involved setting $p=100$. 

In this section, we consider uniform priors on the static state space model parameters, corresponding to hyper-parameter settings specified according to $\beta_{\chi} \sim U\left(0,10\right)$,  $\sigma^2_{\chi} \sim U(0,10)$, $\mu_{\zeta} \sim N\left(0,10\right)$, $\kappa_{\zeta} \sim U\left(0,10\right)$ and $\sigma^2_{\zeta} \sim U(0,10)$. Additionally, we assume the covariance matrix of the additive multivariate Gaussian observation noise on the log of the futures contracts over different maturities, on any given day $t$, is diagonal. We then consider the observation noise variance for the $T_i$-th futures contract on day $t$, denoted $\sigma^2_{\epsilon_i}$ utilised to generate the observations as given by $\Sigma_F =4\mathbb{I}$ which is corresponding to a coefficient of variation in the noise of around $20\%$. Note, in the Bayesian model we treat this observation noise variances as unknown quantities \textit{a priori} that must be estimated jointly with the other model parameters from the posterior. 

We consider a panel of $N$ different futures contracts with maturities $T_1,\ldots,T_N$ and futures prices at time $t$ denoted by $F_{t,T_1},\ldots,F_{t,T_N}$. Note, the filtering framework is set up such that the latent process is the real process with real parameters used and the observation process is the real process with risk neutral parameters used. In the following section we analyse the ability to perform jointly parameter estimation (calibration) and filtering for the latent states of the model as we vary the number of contracts observed on the futures curve on any given day from $N \in \left\{1,5,10\right\}$. 

\begin{figure}[!ht]
\includegraphics[width = 1\textwidth, height = 8cm]{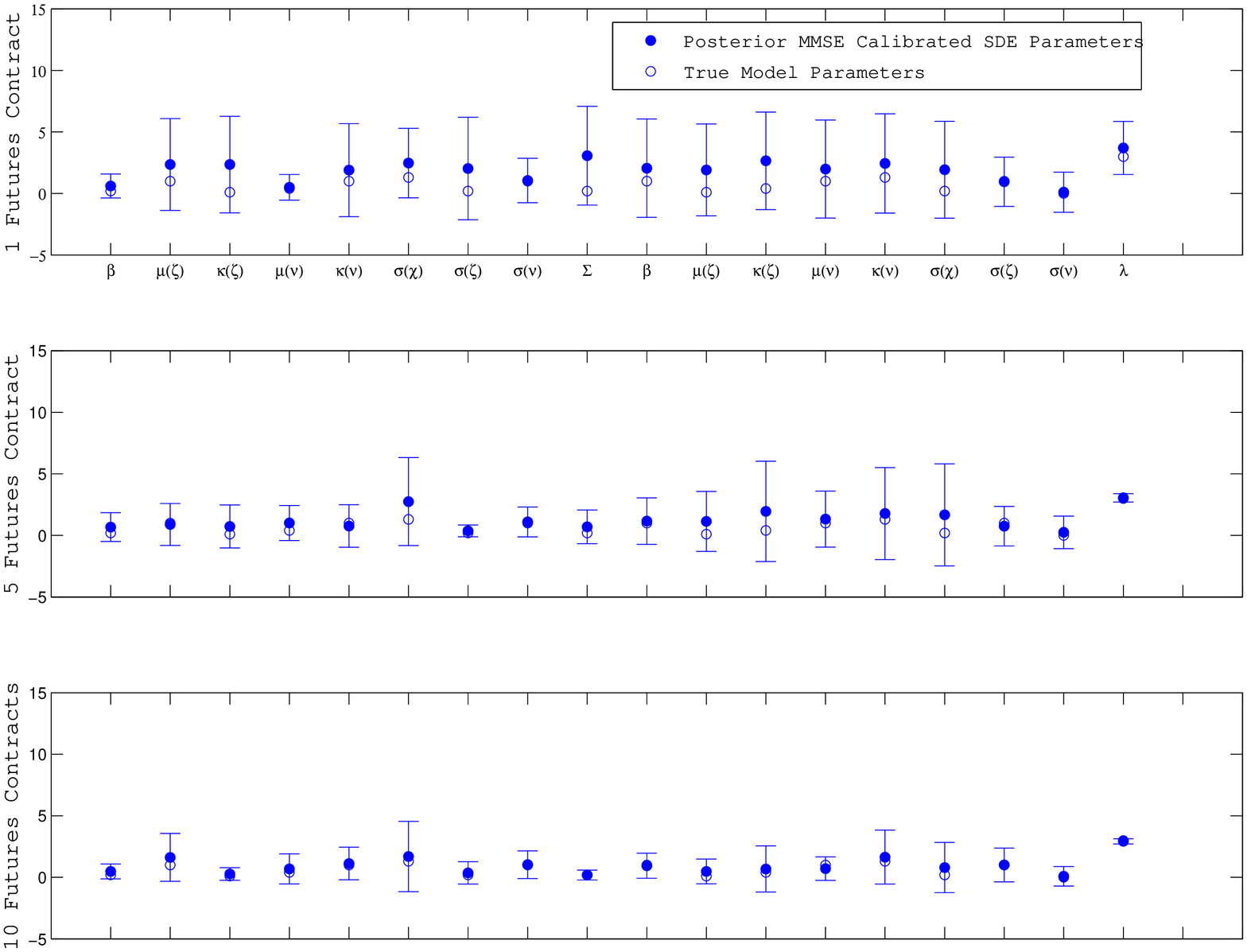}
\caption{Filled circles are the estimated parameter MMSE and error bars represent $95\%$ posterior confidence intervals. Open circles are true parameter values used to generate the data. Parameters are presented in order of real world (left) and risk neutral (right). }
\label{FigMMSE}
\end{figure}

In the following simulation results we demonstrate the accuracy of the joint calibration (parameter estimation) and filtering (state estimation) using the posterior distribution derived in Equation \ref{eqn:Post} and the Rao-Blackwellised Adaptive Particle MCMC algorithm. The first results presented are for the parameter estimations as a function of the number of Futures contracts considered, where we consider $N \in \left\{1,5,10\right\}$. In Figure \ref{FigMMSE} we present the MMSE and $95\%$ posterior confidence intervals for each model parameters real and risk neutral. Each plot contains the parameter estimates as well as the true parameter values used to generate the data represented by circles. The top panel contains the posterior estimates obtained for a panel of 100 days of data with $N=1$ futures contract, the middle panel has posterior estimates for a panel of 100 days of data with $N=5$ futures contracts and the lower panel has these results for 100 days of observations with $N=10$ contracts. Each contract considered is separated by 30 days so we have 30 day contract, 60 day, through to maturity $T_10$ which is a 300 day contract. The results demonstrate that clearly the parameter estimation accuracy is improving as we add more contracts to the panel of data, with $N=10$ contracts having very accurate parameter estimates.

\begin{figure}[!ht]
\includegraphics[width = 0.49\textwidth, height = 5.5cm]{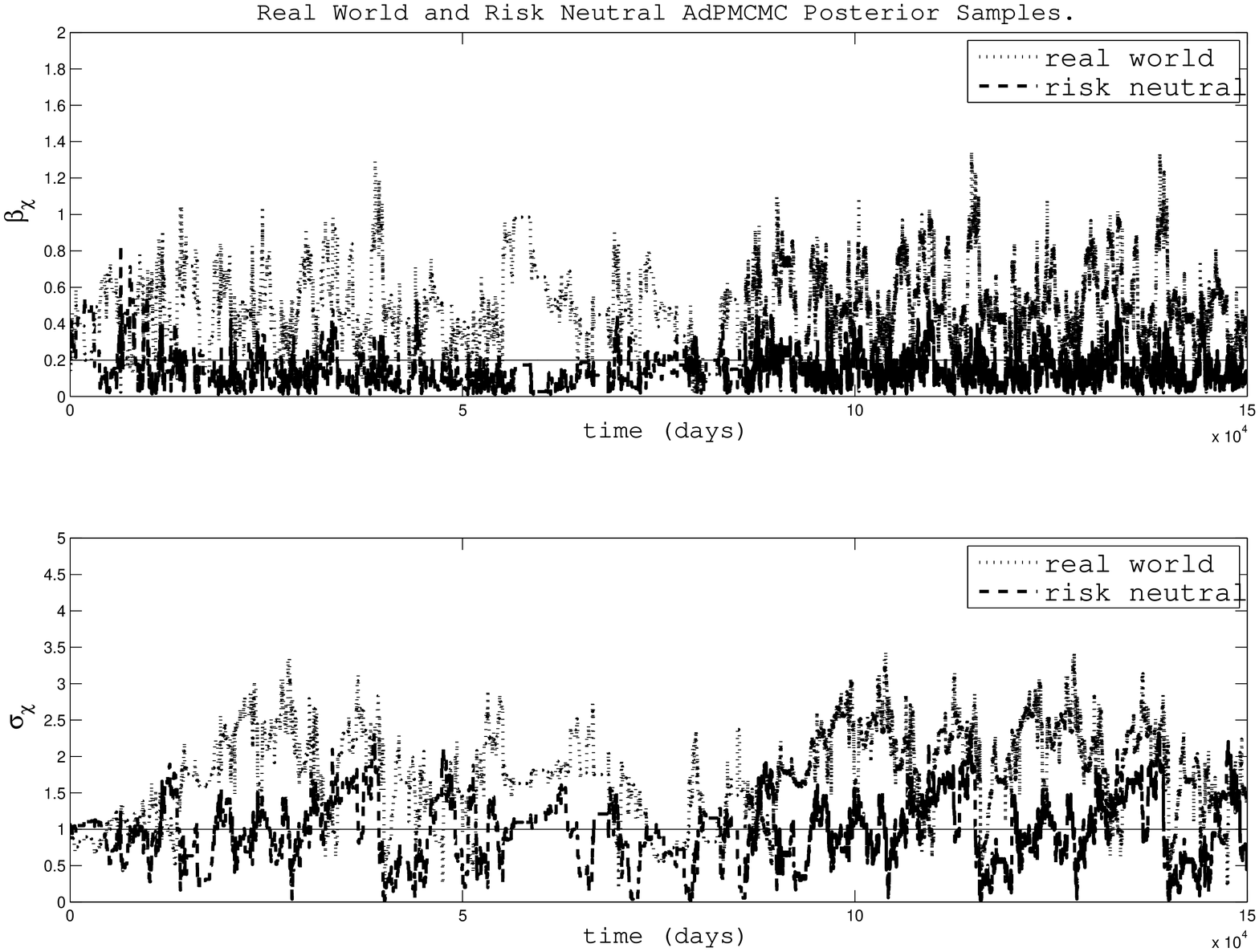}
\includegraphics[width = 0.49\textwidth, height = 5.5cm]{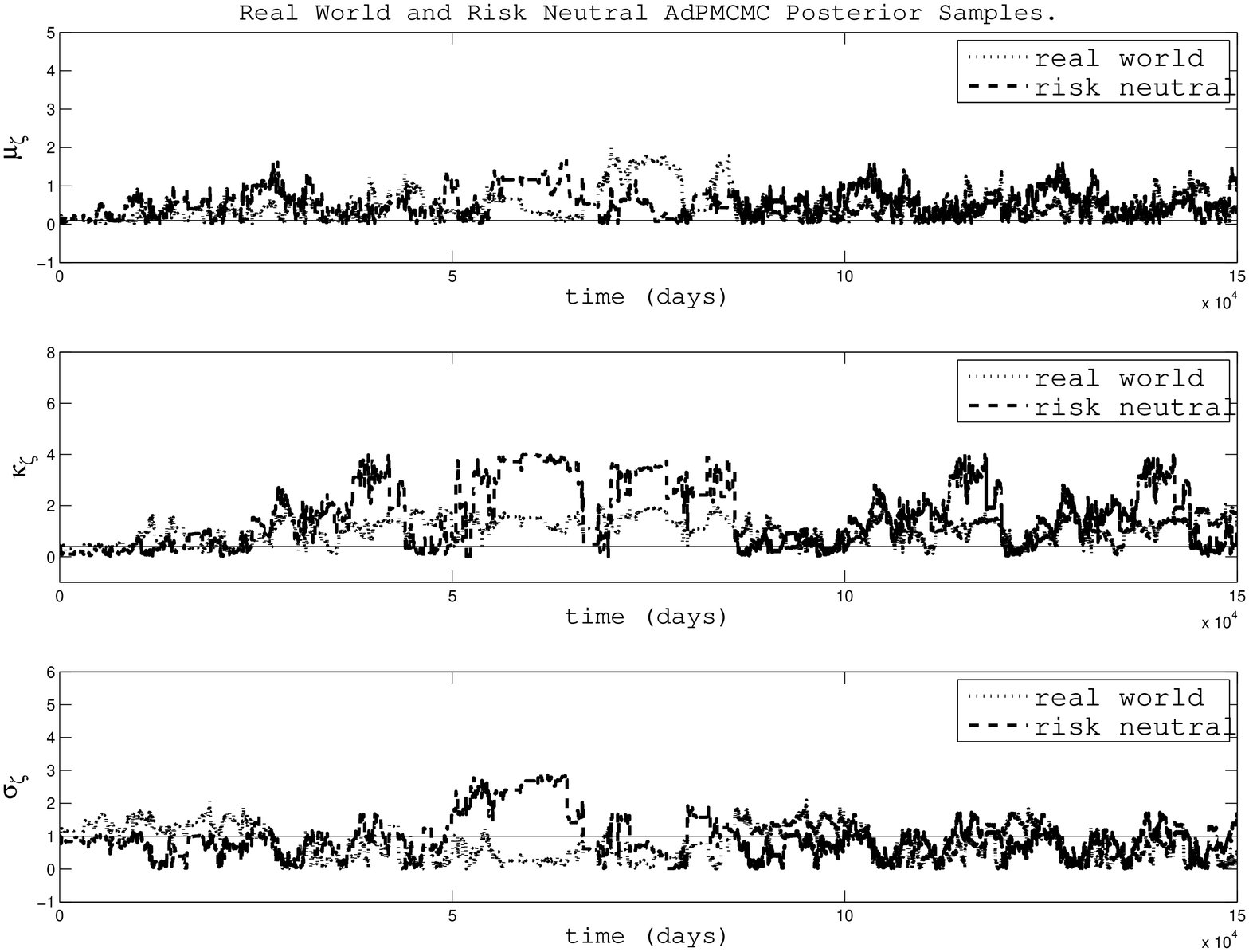}
\includegraphics[width = 0.49\textwidth, height = 5.5cm]{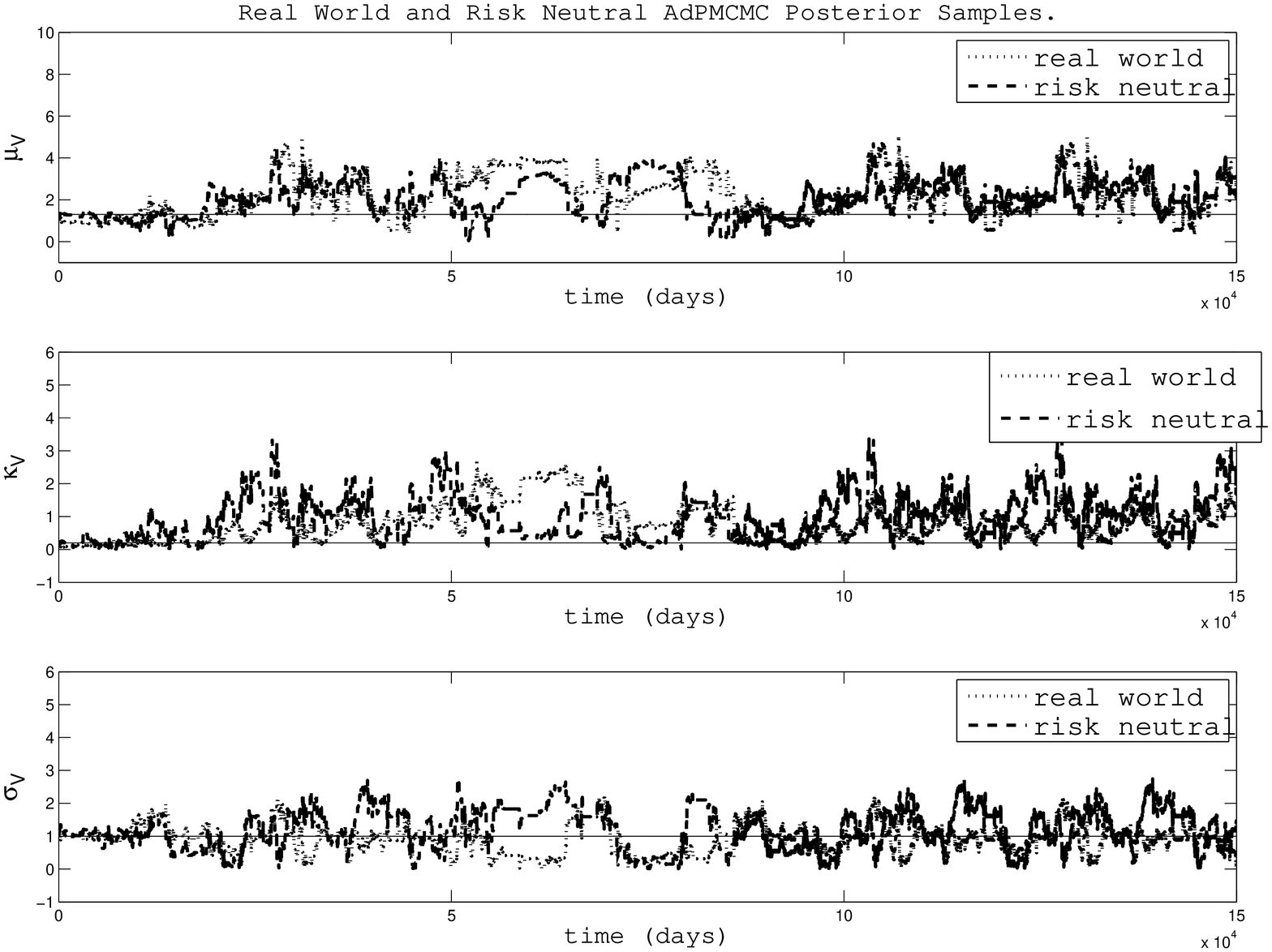}
\includegraphics[width = 0.49\textwidth, height = 5.5cm]{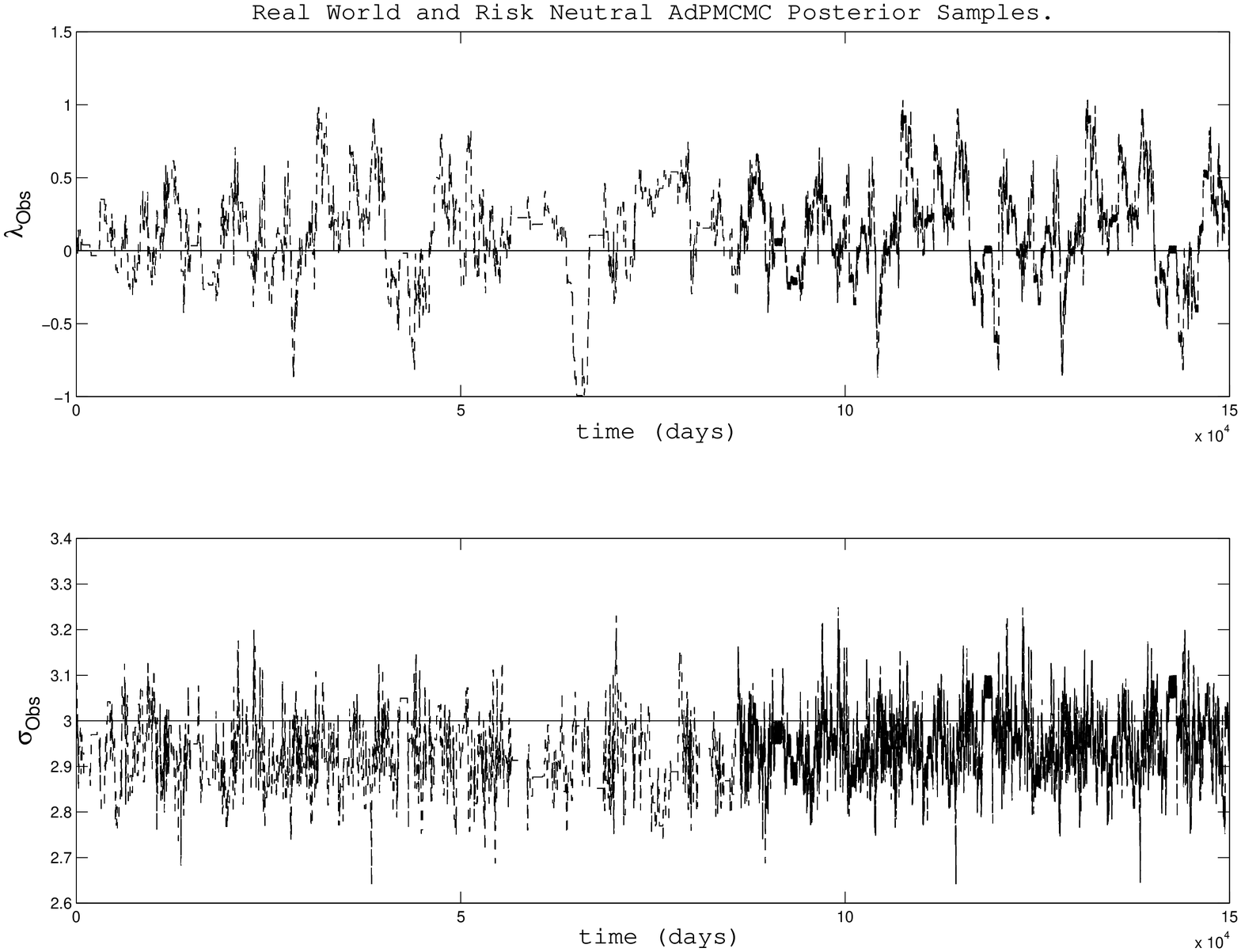}
\caption{Trace plots of the simulated Markov chain state trajectories generated by the Rao-Blackwellised AdPMCMC sampler for the model parameters of the multi-factor s.d.e. model.}
\label{FigTrace}
\end{figure}

In Figure \ref{FigTrace} we present the trace plots of the model parameters from the Rao-Blackwellised Adaptive PMCMC sampler for futures contract panel data contains $N=10$ contracts separated by 30 day maturities over 100 trading days. We have separated the results per factor and present the real world and risk neutral parameters on the same trace plot. These results are for uninformative priors and we see reasonable mixing for the Rao-Blackwellised AdPMCMC sampler updating in dimensions $4T + 18$ each iteration.

In Figure \ref{FigDynamicFactors} we present the Kalman filtered and particle filtered estimates of the latent path space for each of the four factors for short and long term dynamics and the two stochastic volatility terms, for the case of $N=10$ contracts separated by 30 day maturities over 100 trading days in the panel of futures curves. The results demonstrate that the short term factor and volatility terms are very accurately estimated. The long term factor estimates the mean reversion level well, and includes the true latent process within a 95\% posterior confidence interval. 

Finally, the most important result in this analysis is a comparison of the estimated log spot price versus the estimated MMSE log spot price and 95\% posterior confidence interval, see Figure \ref{FigDynamicSpot}. As above, this example corresponds to the situation in which the futures contract panel data contains 10 contracts separated by 30 day maturities over 100 trading days of observations. The Milstein SIR proposal considered in the Rao-Blackwellized Adaptive PMCMC algorithm utilised truncation of $p=100$ and 500 particles.

We also note that simulation studies were also performed utilising futures contracts generated with seasonal components, the resulting estimation accuracy were equivalent to those presented. Additionally, we varied the signal to noise ratio by decreasing observation error to $\sigma_F=1$ and observed a corresponding increase in accuracy of the estimation.   

\subsubsection{Actual Data - Oil Commodity Futures}
\begin{figure}[!ht]
\includegraphics[width = \textwidth, height = 5cm]{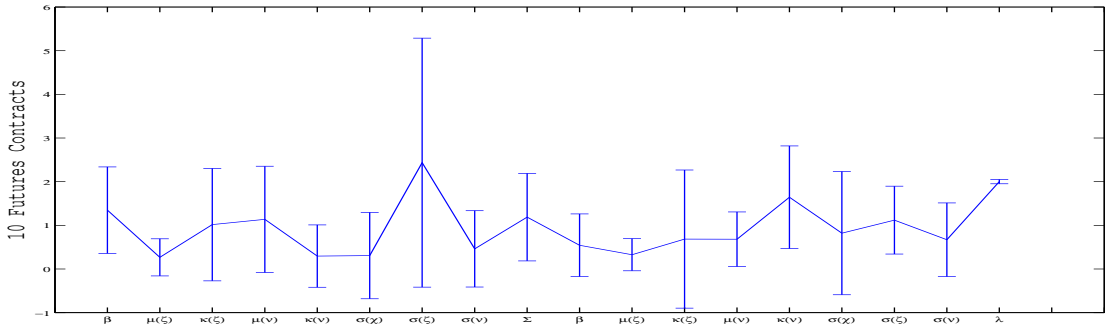}
\caption{Calibrated MMSE posterior model parameters and posterior confidence intervals. This example corresponds to the situation in which the crude oil futures contract panel data contains 10 contracts separated by 20 day maturities over 150 trading days. The Milstein SIR proposal utilised $p=100$ and 500 particles.}
\label{FigOilCalibrations}
\end{figure}

In this section we take a panel 10 futures contracts with daily price data corresponding to the contracts for crude light oil commodity futures over 150 trading days from 2/01/1998 to 6/08/1998. The maturities of the 10 contracts are 21 days for the first contracts initiation and each contract there after has a maturity increasing by 20 days. Therefore the last contract has a maturity from its initiation of  200 days. We estimated the posterior model MMSE parameters and posterior confidence intervals for the developed in Section \ref{BayesModel} using this panel of oil futures data. This was achieved via the Rao-Blackwellised AdPMCMC algorithm for the four factor model developed, again with a calibration and filtering formulation and the 100,000 iterations of the sampler. The average acceptance rate of the PMCMC sampler developed was 28\% and 500 particles were utilised. The Milstein SIR proposal utilised $p=100$ and the calibrated model parameters obtained are presented in Figure \ref{FigOilCalibrations} along with 3 posterior standard deviations of these estimated parameters using the model developed in Section \ref{BayesModel}.

\begin{figure}[!ht]
\includegraphics[width = 0.45\textwidth, height = 6cm]{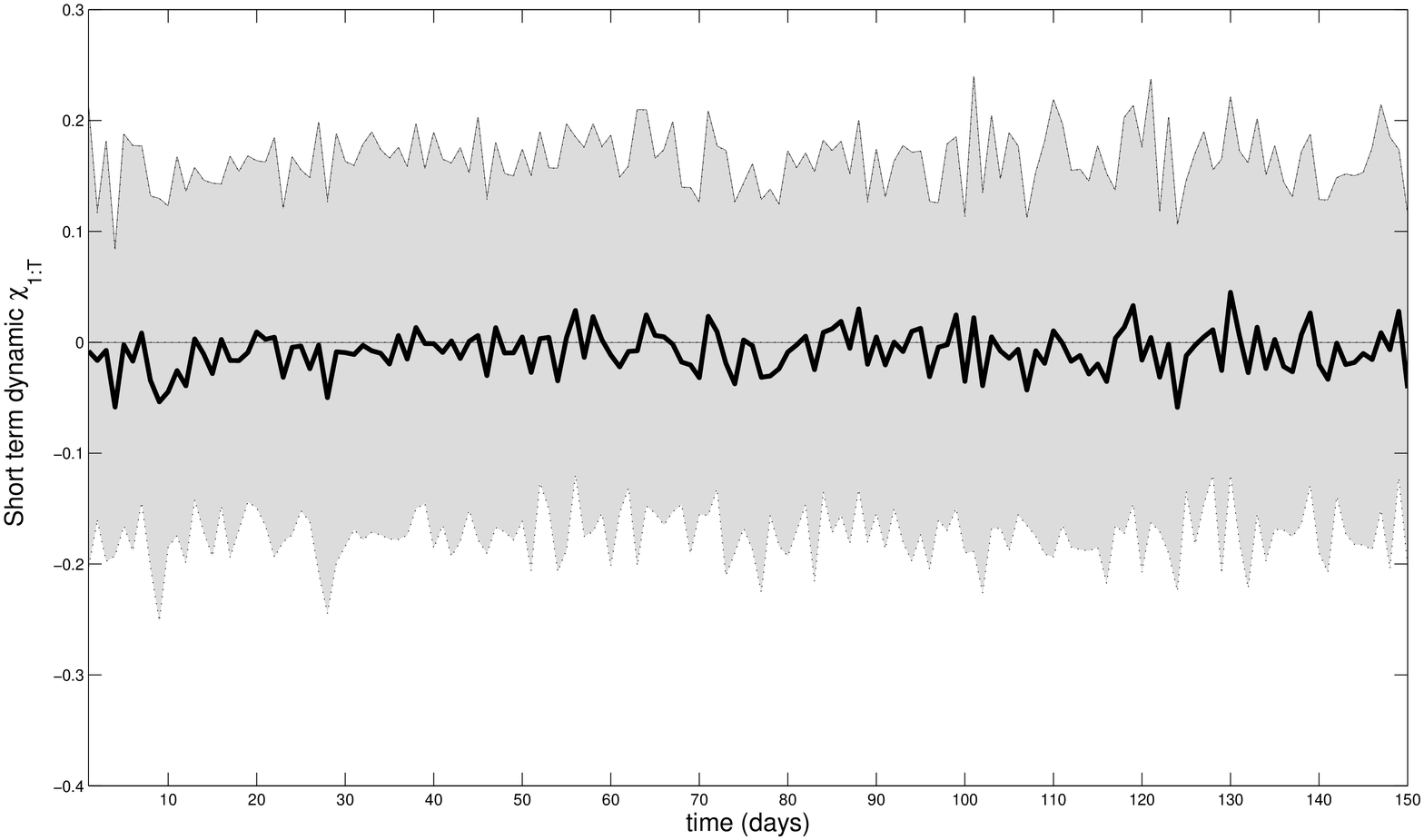}
\includegraphics[width = 0.45\textwidth, height = 6cm]{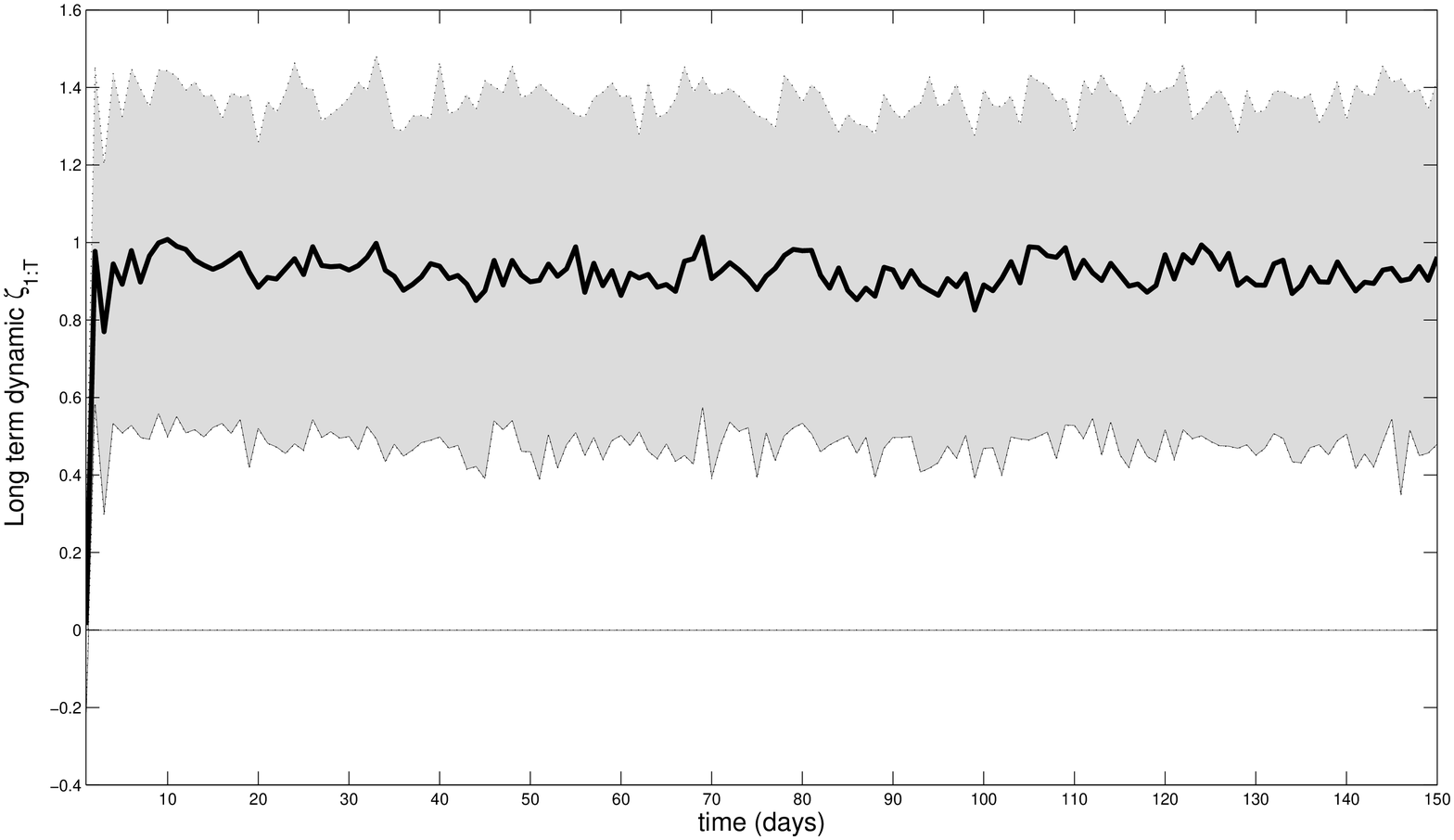}\\
\includegraphics[width = 0.45\textwidth, height = 6cm]{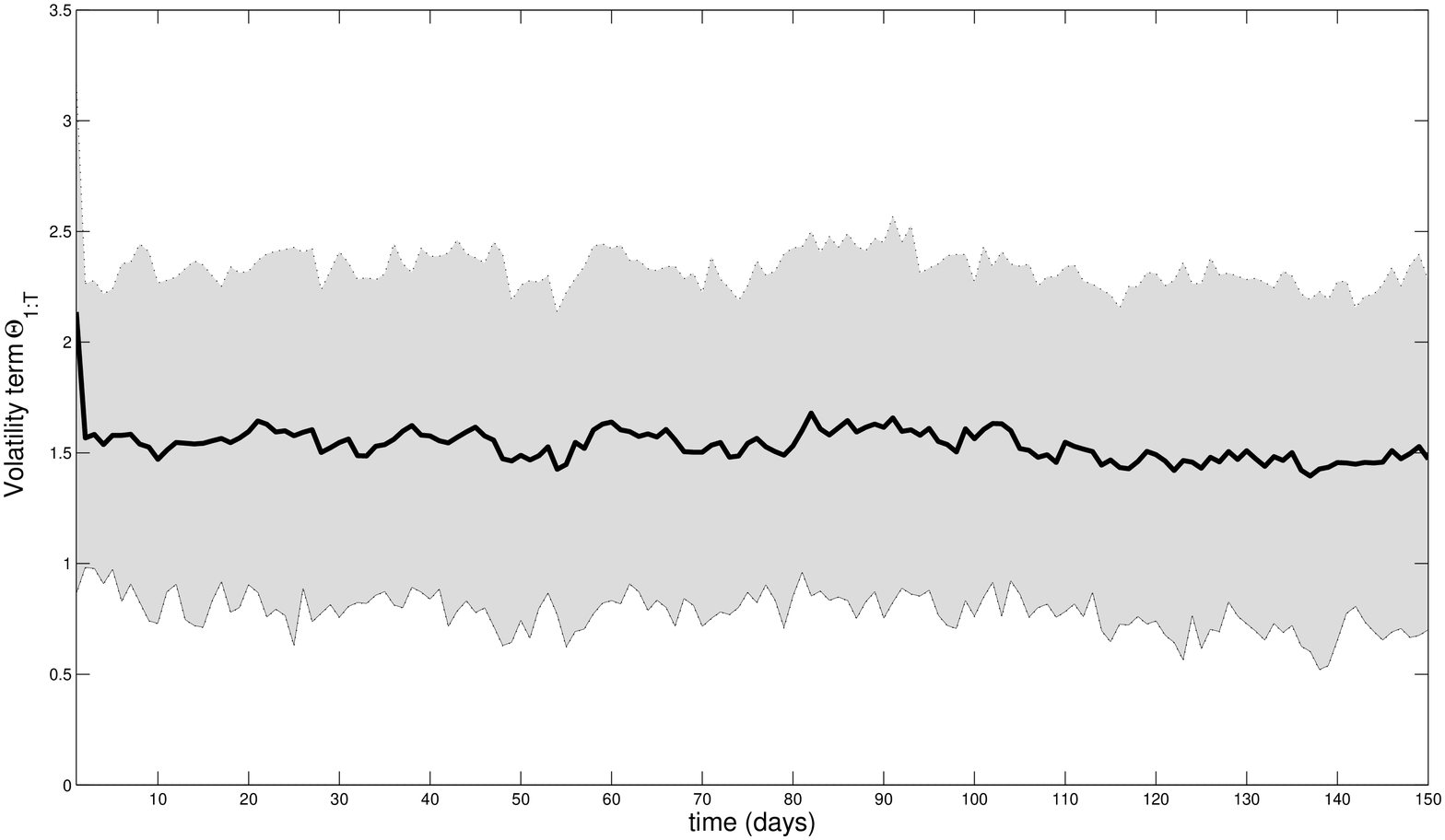}
\includegraphics[width = 0.45\textwidth, height = 6cm]{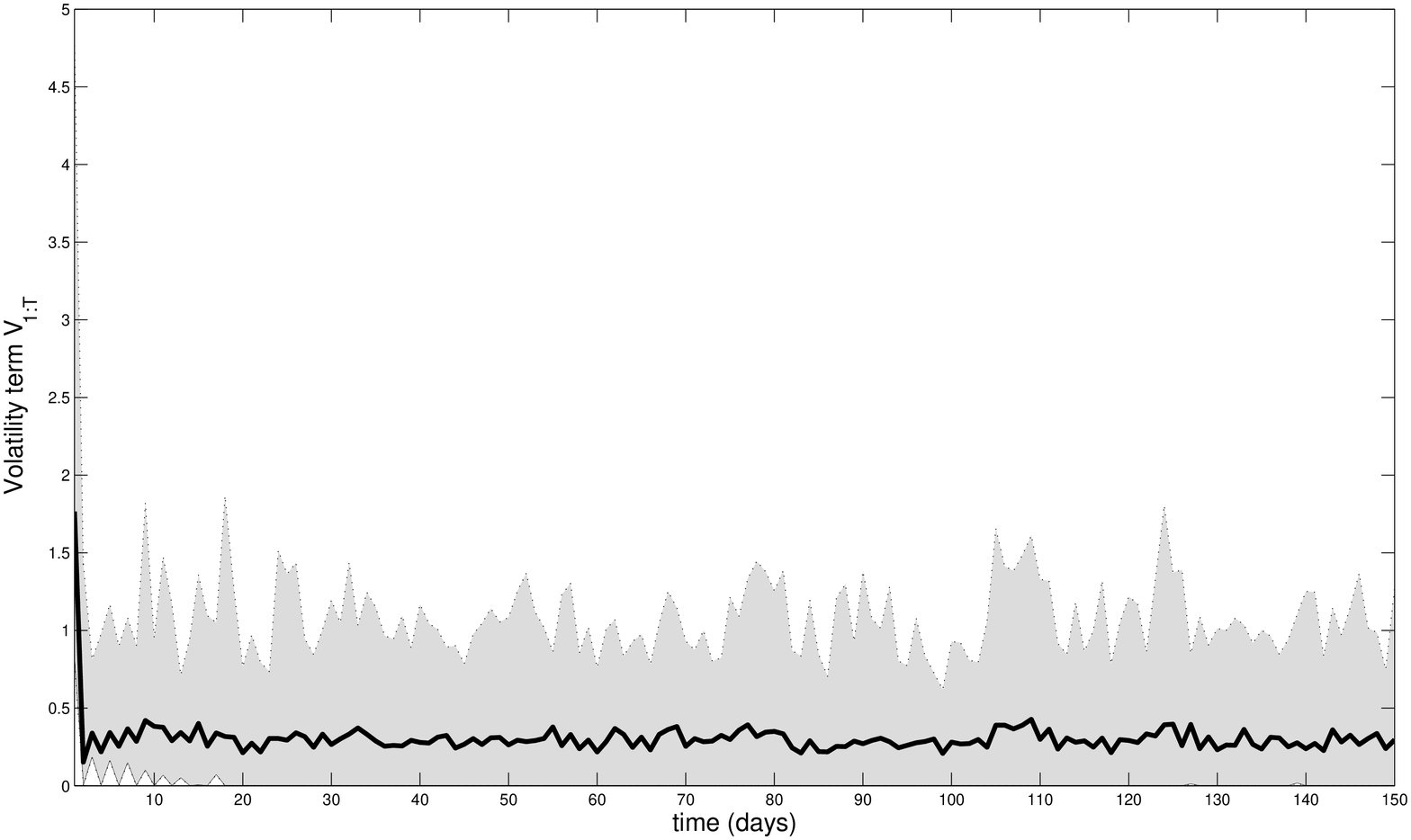}\\
\includegraphics[width = 0.9\textwidth, height = 6cm]{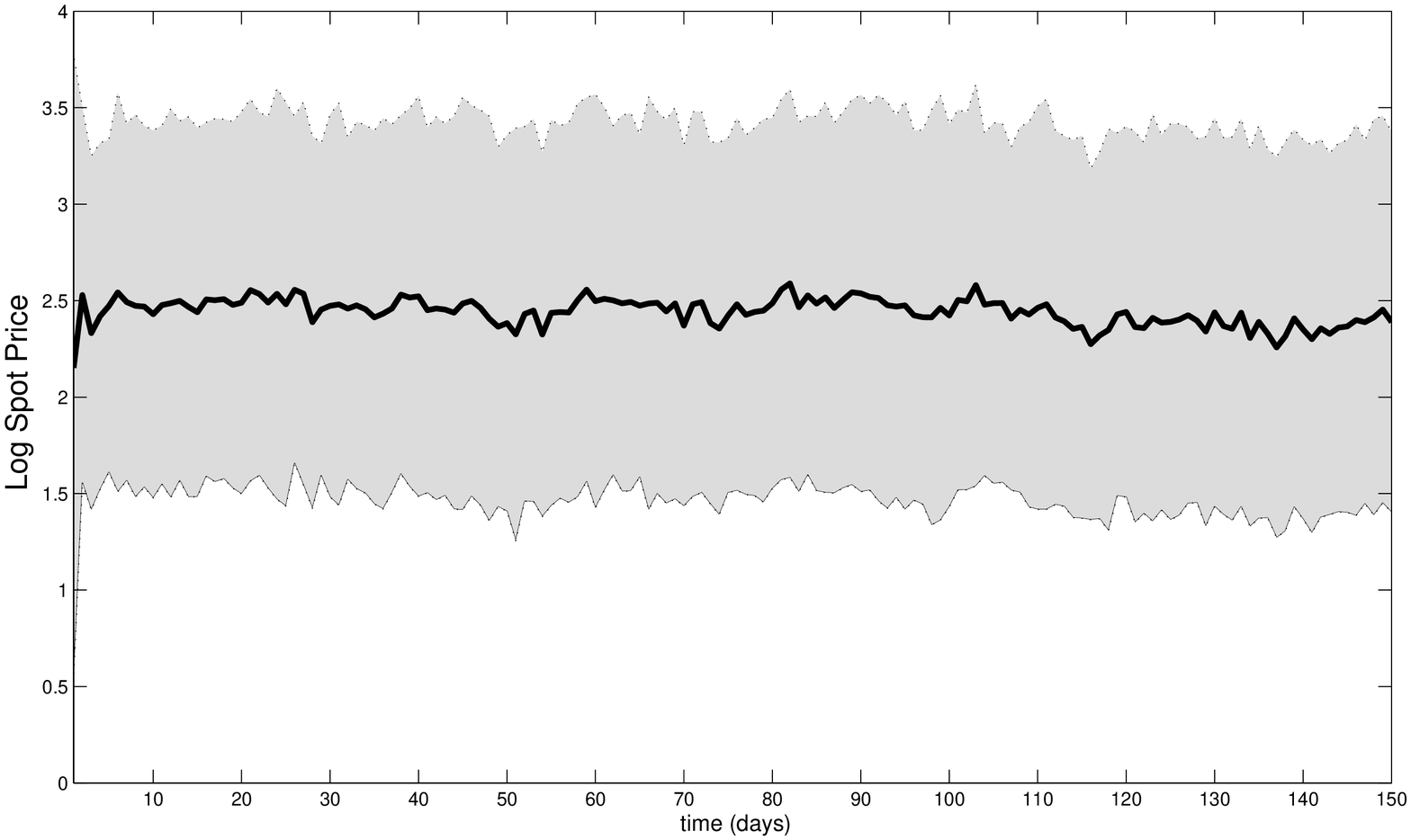}
\caption{Panel 1: Short term dynamic $\chi_{1:T}$, Panel 2: Long term dynamic $\zeta_{1:T}$, Panel 3: Volatility term $\theta_{1:T}$, Panel 4: Volatility $V_{1:T}$, Panel 5: Log spot price estimated $S_{1:T}$. Solid line is filtered MMSE and gray band is posterior 95\% C.I.}
\label{FigOilDynamicSpot}
\end{figure}

In Figure \ref{FigOilDynamicSpot} we present the latent path space trajectory estimates for the four factors over 150 days fitted to the panel of oil futures data. The plots contain the posterior MMSE estimates for each factor and the 95\% posterior C.I. estimates. In addition, we present the estiamted log spot price for the crude oil futures contract data, based on 10 contracts with 150 days of data in the panel. 

\begin{figure}[!ht]
\includegraphics[width = \textwidth, height = 10cm]{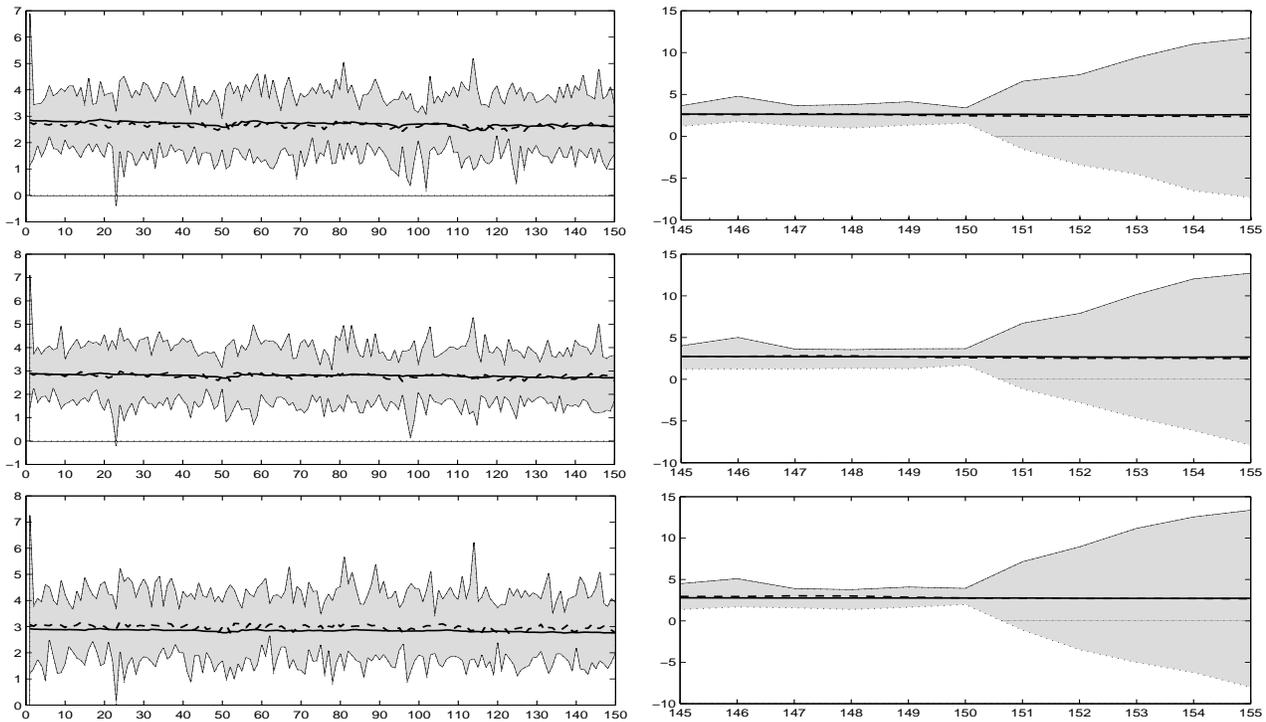}
\caption{Left Panels: within sample prediction, Right Panels: out of sample forecasts. Top Row: contracts with 20 days maturity, Middle Row: contracts with 100 days maturity, Bottom Row: contracts with 200 days maturity. This example corresponds to the situation in which the crude oil futures contract panel data contains 10 contracts separated by 20 day maturities over 150 trading days. Solid line observed log futures price, dashed line predicted MMSE log futures price, gray shading 95\% posterior C.I.}
\label{FigDynamicFuturePred}
\end{figure}

Next we present the MMSEP and the associated posterior predictive 95\% confidence intervals obtained from the predictive distribution of the log futures curves at some time point $\tau$ defined by Equation \ref{MSEPFutures}. 
{\small{
\begin{equation}
\begin{split}
p(&F_{\tau,T_1},F_{\tau,T_2},\ldots,F_{\tau,T_N}|F_{1:T,T_1},F_{1:T,T_2},\ldots,F_{1:T,T_N}) \\
&=  \int \prod_{n=1}^Ng_{\bm{\Phi}}(F_{\tau,T_n}|\chi_{\tau},\zeta_{\tau},\theta_{\tau}) \pi\left(\bm{\Phi},\chi_{1:T},\zeta_{1:T},\theta_{1:T},V_{1:T}|F_{1:T,T_1},F_{1:T,T_2},\ldots,F_{1:T,T_N}\right)d\bm{\Phi}d\chi_{1:T,\tau}d\zeta_{1:T,\tau}d\theta_{1:T,\tau}dV_{1:T,\tau}
\end{split}
\label{MSEPFutures}
\end{equation}
}}
Results in Figure \ref{FigDynamicFuturePred} represent from top to bottom in the figure panels on the left the within sample and on the right the out of sample forecast log futures prices after integrating out uncertainty in the latent process states and parameters, resulting from the posterior predictive distribution. The shaded gray area corresponds to the predicted 95\% posterior predictive confidence interval for the log of the futures prices for contracts 1, 5, 10 observed over 150 days and predicted out of sample over an additional 5 days. The first contract in the top row of results is the contract closest to maturing, the 5th contract in the middle row of results is contracts maturing at 100 days and the 10-th contract in the bottom row of results are those with maturties of 200 days. The solid line represents the observed time series of daily log futures price at close of market over 150 days within sample and 5 days out of sample. The dashed line represents the estimated posterior MMSE and predicted MMSE of the log futures price. Clearly we see that the observed futures price curves are contained within the 95\% posterior C.I of the estimated model at all times, and secondly that the estimated MMSE and forecast MMSE out of sample indicate acturate calibration and estimation of the commodity model on the real crude oil futures data.

\section{Conclusions}
In this paper we have developed a novel general multi-factor model for commodity spot prices and futures valuation. In particular we extend the multi-factor model long-short stochastic differential equation (s.d.e.) model developed in \cite{Schwartz98} and \cite{yan2002valuation} in two important aspects: firstly we allow for both the long and short term dynamic factors to be mean reverting and secondly we develop an additive structural seasonality model. We developed a non-trivial state space model representation for such a model and in doing so we also extend the literature in this area by allowing for futures and options contracts in the observation equation of the Milstein discretized non-linear stochastic volatility state space model. 

Next to perform estimation and calibration under this model we derived a closed form solution for the futures prices and then developed a novel numerical methodology based on Sequential Monte Carlo to perform calibration of the model jointly with the filtering of the latent processes for the long-short and volatility dynamics. In this regard we explore and develop a novel methodology based on an adaptive Rao-Blackwellised version of the Particle Markov chain Monte Carlo methodology for the joint calibration and filtering. In doing this we deal accurately with the non-linearities introduced into the filtering framework. We perform analysis on real data for oil commodities.

\bibliographystyle{IEEEtran}
\bibliography{References}

\section{Appendix 1: Proof of Futures Price} \label{Futures.d.e.rivation}
\textbf{Proof:} Here we present the proof of Theorem 1 for the futures price from the real latent process s.d.e. model.
To proceed we obtain the risk neutral process from the real process we need to deduct the risk free premium from the long term trend to account for the market price of risk where the risk premium is denoted by $\lambda\left(x,t\right)$. Next, we denote the Futures contract price at time $t$ with maturity of the contract at $T$ by $F(t,T)$.
\begin{align*}
F(t,T) &= \mathbb{E}^{\ast}\left[\exp\left(X(T)\right)|X(t)\right]
=\mathbb{E}^{\ast}\left[\exp\left(f(T)+\xi^{\ast}_{T}+\chi^{\ast}_{T}+\theta^{\ast}_{T}\right)|f(t),\xi^{\ast}_{t},\chi^{\ast}_{t},\theta^{\ast}_{t}\right],
\end{align*}
where $\mathbb{E}^{\ast}$ denotes the expectation taken with respect to the risk neutral process. To solve for this expectation, we must first derive the expression for the transition density of the state variables, denoted by
$p\left(\xi^{\ast}_{T}+\chi^{\ast}_{T}+\theta^{\ast}_{T},
V_T^{\ast},T|\xi^{\ast}_{t},\chi^{\ast}_{t},\theta^{\ast}_{t},V_t^{\ast},t\right)=p\left(\bm{Y}_{T}|\bm{Y}_{t},t\right)$.
This is obtained for the continuous s.d.e. latent process model by writing down the either the Kolmogorov-backward or forward equation. Here we work with the backward equation (dropping the $\ast$ notation on state variables for convenience) with the condition that, $p\left(\bm{Y}_{T},T|\bm{Y}_{t},t=T\right)=\delta(\bm{Y}_T-\bm{Y}_t)$. Next, to obtain the futures price we multiply each term in the Kolmogorov-backward equation by $\exp\left(X_t\right)$ and then
integrate with respect to $X_t$. Note, the notation $F_{t,T}$ refers to $F(\bm{Y}_t,t=T,T)$. This allows us to express each term in the pde with respect to the futures price. 
{\small{
\begin{align*}
&\frac{\partial F_{t,T}}{\partial t} +
\frac{1}{2}\sigma_{\xi}^2\frac{\partial^2 F_{t,T}}{\partial
\xi_{t}^2} + \frac{1}{2}\sigma_{\chi}^2\frac{\partial^2
F_{t,T}}{\partial \chi_{t}^2} +
\frac{1}{2}\sigma_{V}^2V_t\frac{\partial^2 F_{t,T}}{\partial
V_{t}^2} + \frac{1}{2}V_t\frac{\partial^2 F_{t,T}}{\partial
\theta_{t}^2} \\
&+ \left(\mu_{\xi}^{\ast}-\kappa_{\xi}^{\ast}\xi_{t}\right)
\frac{\partial F_{t,T}}{\partial \xi_{t}} +
\left(\beta^{\ast}\chi_t-\lambda_0\right) \frac{\partial
F_{t,T}}{\partial \chi_{t}} +
\left(\mu_{V}^{\ast}-\kappa_{V}^{\ast}V_{t}\right) \frac{\partial
F_{t,T}}{\partial V_{t}} + \left(-\lambda_4 V_t\right)
\frac{\partial F_{t,T}}{\partial
\theta_{t}} \\
&+ \rho_{\chi \xi}\sigma_{\chi}\sigma_{\xi} \frac{\partial^2
F_{t,T}}{\partial \chi_{t} \partial \xi_{t}} + \rho_{\theta
V}\sigma_{V}V_t \frac{\partial^2 F_{t,T}}{\partial \chi_{t}
\partial V_{t}} = 0,
\end{align*}
}} with condition $F(\bm{Y}_t,t=T,T) = \exp(X_T)$.

Next, since the state variables enter into the above pde in a linear fashion, therefore the solution belongs to
the class of exponentially affine models. Hence, we can look for a solution of the form,
\begin{equation*}
F_{t,T} = \exp\left(B_0(\tau) + B_1(\tau)\xi_t + B_2(\tau)\chi_t +
B_3(\tau)\theta_t + B_4(\tau)V_t \right).
\end{equation*}
Hence, substitution of this solution results in a system of 5 ODE's which after substitution and grouping of like terms, then divide both sides by $F_{t,T}$ and group terms according to
latent factors to obtain: 
{\small{
\begin{align*}
&\frac{dB_1(\tau)}{d\tau} = - \kappa_{\xi}^{\ast} B_1(\tau);\\
&\frac{dB_2(\tau)}{d\tau} = - \beta^{\ast} B_2(\tau);\\
&\frac{dB_3(\tau)}{d\tau} = 0;\\
&\frac{dB_4(\tau)}{d\tau} = - \lambda_4 B_3(\tau) -
\kappa_V^{\ast}B_4(\tau) +
\frac{1}{2}\sigma_V^2\left[B_4(\tau)\right]^2 + \rho_{\theta
V}\sigma_{V} B_3(\tau)B_4(\tau) + \frac{1}{2}B^2_3(\tau);\\
&\frac{dB_0(\tau)}{d\tau} =  -\frac{1}{2}\sigma_{\xi}^2
\left[B_1(\tau)\right]^2 - \frac{1}{2}\sigma_{\chi}^2
\left[B_2(\tau)\right]^2 - B_1(\tau)\mu_{\xi}^{\ast} +
\lambda_0B_2(\tau) - \mu_{V}^{\ast}B_4(\tau) - \rho_{\chi
\xi}\sigma_{\chi}\sigma_{\xi} B_2(\tau)B_1(\tau)
\end{align*}
}} 
Note, the condition that $F(Y_t,t=T,T) = \exp(X_t)$ needs to be
satisfied, in addition we assumed the form of the risk premium
$\lambda(X_t,t) = \lambda_4 \sqrt{V_t} = -\frac{1}{2}\sqrt{V_t}$
(this allows us to remove the volatility $V_t$ from the futures
price, however still keep it in the options price).

The solution to this system of ODE's is obtained by first applying the constraint at time $T$ given by $F(Y_t,t=T,T) =
\exp(X_T)$ which automatically sets $B_1(0)=B_2(0)=B_3(0)=1$ and
$B_0(0)=B_4(0)=0$. We can then obtain solutions $B_2(\tau)=\exp(-\beta^{\ast}\tau)$, $B_1(\tau)=\exp(-\kappa_{\xi}^{\ast}\tau)$ and with initial condition $B_3(0) = 1$ we get $B_3(\tau)=1$. Finally, substituting the solution for $B_3(\tau)$ into
the expression for $\frac{dB_4(\tau)}{d\tau}$ and setting
$\lambda_4 = -\frac{1}{2}$ results in ODE
$$\frac{dB_4(\tau)}{d\tau} = \left[\rho_{\theta V}\sigma_{V} -
\kappa_V^{\ast} \right]B_4(\tau) +
\frac{1}{2}\sigma_V^2\left[B_4(\tau)\right]^2$$ with the initial condition that $B_4(0)=0$. This is recognized as Ricatti's equation of the 2nd kind one can obtain a solution to a transformed linear 2nd order ode and then in turn obtain a solution to the Ricatti equation which produces a solution $B_4(\tau) = 0$.
Finally, substitution of these solutions for
$B_1(\tau),B_2(\tau)$ and $B_4(\tau)$ into the expression for
$\frac{dB_0(\tau)}{d\tau}$ results in an ode which can be integrated (w.r.t. $\tau$) explicitly to obtain a solution for $B_0(\tau)$ as follows:
\begin{align*}
B_0(\tau) &= -\frac{1}{2}\sigma_{\xi}^2
\int\exp(-2\kappa_{\xi}^{\ast}\tau)d\tau -
\frac{1}{2}\sigma_{\chi}^2 \int\exp(-2\beta^{\ast}\tau)d\tau -\\
&\int \exp(-\kappa_{\xi}^{\ast}\tau)\mu_{\xi}^{\ast}d\tau +
\lambda_0\int\exp(-\beta^{\ast}\tau)d\tau -
\rho_{\chi\xi}\sigma_{\chi}\sigma_{\xi}
\int\exp(-\beta^{\ast}\tau)\exp(-\kappa_{\xi}^{\ast}\tau)d\tau \\
&= \frac{\sigma_{\xi}^2}{4\kappa_{\xi}^{\ast}}
\exp(-2\kappa_{\xi}^{\ast}\tau) +
\frac{\sigma_{\chi}^2}{4\beta^{\ast}}\exp(-2\beta^{\ast}\tau) +\\
&\frac{\mu_{\xi}^{\ast}}{\kappa_{\xi}^{\ast}}
\exp(-\kappa_{\xi}^{\ast}\tau) -
\frac{\lambda_0}{\beta^{\ast}}\exp(-\beta^{\ast}\tau) +
\frac{\rho_{\chi\xi}\sigma_{\chi}\sigma_{\xi}}{\beta^{\ast} +
\kappa_{\xi}^{\ast}}\exp(-\beta^{\ast}\tau -
\kappa_{\xi}^{\ast}\tau)
\end{align*}
\qed

\section{Appendix 2: Proof of Strong Stochastic Taylor Representation of s.d.e. Factors} \label{BivariateMilstein}
The strong Taylor stochastic expansion known as the Milstein scheme in a multivariate setting is presented below. We present
the i-th component of a general n-dimensional s.d.e., with m-dimensional Weiner process as,
\begin{align*}
&dX_t^i = a^i\left(t,X_t\right)dt +
\sum^m_{j=1}b^{i,j}\left(t,X_t\right)dW_t^j.
\end{align*}
Throughout the remainder of this section we will adopt here the operator notation of Kloeden and Platen (1999). In their book, p.
346 they demonstrate that the i-th component of the Milstein scheme under the Ito integrals is then given by,
\begin{align*}
X_t^i &= X_{t-1}^i + a^i\left(t-1,X_{t-1}\right)\triangle t +
\sum_{j=1}^m b^{i,j}\left(t-1,X_{t-1}\right)\triangle W_{t-1}^j \\
& + \sum_{j_1,j_2}^m
L^{j_1}b^{i,j_2}\left(t-1,X_{t-1}\right)I_{(j_1,j_2)\triangle t}
\end{align*}
where, $L^{j}=\sum_{i=1}^n b^{i,j}\frac{\partial}{\partial x_i}$
and the Ito multiple integral $I_{(j_1,j_2)\triangle t}$ is given
by,
\begin{align*}
&I_{(j_1,j_2)\triangle t} = \int_{t_n}^{t_{n+1}}
\int_{t_n}^{s_1}dW_{s_1}^{j_1}dW_{s_1}^{j_2}.
\end{align*}
These integrals have the useful properties that,
\begin{align*}
&I_{(j_1,j_1)\triangle t} = \frac{1}{2}\{\left(\triangle W_t\right)^2 -
\triangle_t \},
\end{align*}
and $I_{(j_1,j_1)\triangle t}$ with $j_1 \neq j_2$ not easily expressed.

As pointed out in Kloeden and Platen (1999), when $j_1 \neq j_2$
the Ito and Stratonovich integrals are equal,
\begin{equation*}
I_{(j_1,j_2)}=J_{(j_1,j_2)}=\int_{\tau_n}^{\tau_{n+1}}\int_{\tau_{n}}^{s_1}dW_{s_2}^{j_1}dW_{s_1}^{j_2}.
\end{equation*}
It will be easier to obtain the approximation under the
Stratonovich integrals which under a p-th order truncation is
given by,
\begin{equation*}
\begin{split}
J_{(j_1,j_2)\triangle t}^p &= \triangle t \left(
\frac{1}{2}\zeta_{j_1}\zeta_{j_2} +
\sqrt{\rho_{p}}\left(\mu_{j_1,p}\zeta_{j_2} -
\mu_{j_2,p}\zeta_{j_1}\right)\right) \\
&+ \frac{\triangle t}{2 \pi} \sum_{r=1}^p
\frac{1}{r}\left(\psi_{j_1,r}\left(\sqrt{2}\zeta_{j_2} +
\nu_{j_2,r}\right) - \psi_{j_2,r}\left(\sqrt{2}\zeta_{j_1} +
\nu_{j_1,r}\right)\right),
\end{split}
\label{eqn:Jmix}
\end{equation*}
where $\zeta_{j},\mu_{j,p},\nu_{j,r}$ and $\psi_{j,r}$ are all
independent $\mathcal{N}(0;1)$ Gaussian random variables with,
\begin{equation*}
\begin{split}
\rho_p &= \frac{1}{12} - \frac{1}{2 \pi^2}\sum_{r=1}^p\frac{1}{r^2},\\
\zeta_j &= \frac{1}{\sqrt{\triangle t}} \triangle W^j.
\end{split}
\end{equation*}

We note that a bivariate approximation for the Heston model was
studied in Stump (...) and a recommendation for p is also provided
in Kloeden and Platen (1999), who suggest $p \geq
\frac{K}{\triangle t}$ for some positive constant $K$.

Hence, the bivariate s.d.e.'s for $\theta_t$ and $V_t$ given by,
\begin{align*}
&d\theta_{t}= \sqrt{V_t}dZ_{\theta} \\
&dV_{t}=\left(\mu_{V} - \kappa_{V}V_{t}\right)dt +
\sigma_{V}\sqrt{V_t}dZ_{V}
\end{align*}
will first be recast with respect to independent Weiner processes $dW_1$ and $dW_2$ as follows
\begin{align*}
&d\theta_{t}= \sqrt{V_t}dW_1 \\
&dV_{t}=\left(\mu_{V} - \kappa_{V}V_{t}\right)dt + 
\sigma_{V}\sqrt{V_t}\left(\rho_{V \theta} dW_1 + \sqrt{1 - \rho^2_{V \theta}} dW_2\right).
\end{align*}

This will result in the following specifications, (n=m=2):
\begin{equation*}
\begin{split}
&a^1\left(t,\theta_t, V_t\right) = 0; a^2\left(t,\theta_t, V_t\right) = \left(\mu_{V} - \kappa_{V}V_{t}\right);
b^{1,1}\left(t,\theta_t, V_t\right)=\sqrt{V_t};
b^{1,2}\left(t,\theta_t, V_t\right) = 0\\
&b^{2,1}\left(t,\theta_t, V_t\right) = \sigma_{V}\sqrt{V_t}\rho_{V \theta};
b^{2,2}\left(t,\theta_t, V_t\right)= \sigma_{V}\sqrt{V_t}\sqrt{1 - \rho^2_{V \theta}}\\
&L^1b^{1,1}\left(t,\theta_t, V_t\right) =
b^{1,1}\left(t,\theta_t, V_t\right)\frac{\partial}{\partial
\theta_t}b^{1,1}\left(t,\theta_t, V_t\right) +
b^{2,1}\left(t,\theta_t, V_t\right)\frac{\partial}{\partial
V_t}b^{1,1}\left(t,\theta_t, V_t\right) = \frac{1}{2}\sigma_{V}\rho_{V \theta}\\
&L^2b^{1,1}\left(t,\theta_t, V_t\right) =
b^{1,2}\left(t,\theta_t, V_t\right)\frac{\partial}{\partial
\theta_t}b^{1,1}\left(t,\theta_t, V_t\right) +
b^{2,2}\left(t,\theta_t, V_t\right)\frac{\partial}{\partial
V_t}b^{1,1}\left(t,\theta_t, V_t\right)
= \frac{1}{2} \sigma_{V} \sqrt{1-\rho^2_{V \theta}}\\
&L^1b^{1,2}\left(t,\theta_t, V_t\right) = 0; L^2b^{1,2}\left(t,\theta_t, V_t\right) = 0;\\
&L^2b^{2,1}\left(t,\theta_t, V_t\right) = b^{1,2}\left(t,\theta_t, V_t\right)\frac{\partial}{\partial
\theta_t}b^{2,1}\left(t,\theta_t, V_t\right) +
b^{2,2}\left(t,\theta_t, V_t\right)\frac{\partial}{\partial
V_t}b^{2,1}\left(t,\theta_t, V_t\right)
= \frac{1}{2} \sigma^2_{V} \rho_{V \theta} \sqrt{1-\rho^2_{V \theta}}\\
&L^1b^{2,2}\left(t,\theta_t, V_t\right) = b^{1,1}\left(t,\theta_t, V_t\right)\frac{\partial}{\partial
\theta_t}b^{2,2}\left(t,\theta_t, V_t\right) +
b^{2,1}\left(t,\theta_t, V_t\right)\frac{\partial}{\partial
V_t}b^{2,2}\left(t,\theta_t, V_t\right) = \frac{1}{2} \sigma^2_{V} \rho_{V \theta} \sqrt{1-\rho^2_{V \theta}}\\
&L^2b^{2,2}\left(t,\theta_t, V_t\right) = b^{1,2}\left(t,\theta_t, V_t\right)\frac{\partial}{\partial
\theta_t}b^{2,2}\left(t,\theta_t, V_t\right) +
b^{2,2}\left(t,\theta_t, V_t\right)\frac{\partial}{\partial
V_t}b^{2,2}\left(t,\theta_t, V_t\right) = \frac{1}{2} \sigma^2_{V} \left(1-\rho^2_{V \theta}\right)\\
&L^1b^{2,1} = b^{1,1}\left(t,\theta_t, V_t\right)\frac{\partial}{\partial
\theta_t}b^{2,1}\left(t,\theta_t, V_t\right) +
b^{2,1}\left(t,\theta_t, V_t\right)\frac{\partial}{\partial
V_t}b^{2,1}\left(t,\theta_t, V_t\right) = \frac{1}{2} \sigma^2_{V} \rho^2_{V \theta}\\
\end{split}
\end{equation*}
This leads to the following bivariate strong Milstein discretization scheme,
\begin{align*}
&\theta_{t}= \theta_{t-1} + \sqrt{V_{t-1}\triangle t}n_{\theta,t} + \frac{1}{2}\sigma_V \rho_{V \theta} I_{(1,1)\triangle t} + \frac{1}{2}\sigma_V \sqrt{1-\rho^2_{V \theta}} I_{(2,1)\triangle t} \\
&\approx \theta_{t-1} + \sqrt{V_{t-1}\triangle t}n_{\theta,t} + \frac{1}{4}\sigma_v \rho_{V \theta} \left(\triangle t n^2_{\theta,t} - \triangle t\right) + \frac{1}{2}\sigma_V \sqrt{1-\rho^2_{V \theta}} J^p_{(2,1)\triangle t}\\
&V_t = V_{t-1} + \left(\mu_{V} - \kappa_{V}V_{t-1}\right) \triangle t + \sigma_V \rho_{V \theta} \sqrt{V_{t-1} \triangle t} n_{\theta,t} + \sigma_V \sqrt{V_{t-1} \left(1-\rho^2_{V \theta}\right) \triangle t} n_{V,t} \\
&+ \frac{1}{2}\sigma_V^2 \rho_{V \theta}^2 I_{(1,1) \triangle t} + 
\frac{1}{2}\sigma_V^2 \rho_{V \theta} \sqrt{1-\rho_{V \theta}^2} I_{(1,2) \triangle t} + 
\frac{1}{2}\sigma_V^2 \rho_{V \theta} \sqrt{1-\rho_{V \theta}^2} I_{(2,1) \triangle t} + 
\frac{1}{2}\sigma_V^2 \left(1-\rho_{V \theta}^2\right) I_{(2,2) \triangle t}\\
&\approx V_{t-1} + \left(\mu_{V} - \kappa_{V}V_{t-1}\right) \triangle t + \sigma_V \rho_{V \theta} \sqrt{V_{t-1} \triangle t} n_{\theta,t} + \sigma_V \sqrt{V_{t-1} \left(1-\rho^2_{V \theta}\right) \triangle t} n_{V,t} \\
&+ \frac{1}{4}\sigma_V^2 \rho_{V \theta}^2 \left(\triangle t n^2_{\theta,t} - \triangle t \right) + 
\frac{1}{2}\sigma_V^2 \rho_{V \theta} \sqrt{1-\rho_{V \theta}^2} J^p_{(1,2) \triangle t} + 
\frac{1}{2}\sigma_V^2 \rho_{V \theta} \sqrt{1-\rho_{V \theta}^2} J^p_{(2,1) \triangle t} + 
\frac{1}{4}\sigma_V^2 \left(1-\rho_{V \theta}^2\right) \left(\triangle t n^2_{V,t} - \triangle t \right)
\end{align*}
where $n_{\theta,t}$ and $n_{V,t}$ are i.i.d. standard normal
random variables and we will approximate the mixed expressions $J_{2,1}(\theta)$ and $J_{1,2}(V)$ with $p$-th order truncations of the series, when we perform simulation of the series.

\section{Appendix 3: Simulating a Strong Stochastic Taylor Scheme} \label{SimulationBivariateMilstein}
Simulation of the strong stochastic Taylor scheme for a Miltstein approximation is give next. This will form an important aspect of the proposal mechanism in the SIR fitler contained in the Particle MCMC algorithm. We wish to simulate new realizations for $\theta_t$ and $V_t$ given $\theta_{t-1}$ and $V_{t-1}$ and model parameters at time t, $\bm{\Phi}$.

\begin{table}[htb]
\centering
\begin{tabular}{llll}
\multicolumn{4}{l}{{\bf Simulation of a bivariate Milstein Scheme.}}\\
\hline
\multicolumn{4}{l}{{\textbf 1. Initialisation:}}\\
    & \multicolumn{2}{l}{\textbf{i.} Specify time discretization $\triangle t$ and truncation $p$.} & \\
\multicolumn{4}{l}{{\textbf 2. Simulation:}}\\    
 & {\it For $i=1,\ldots,N$ }& & \\ 
 & \textbf{i.} simulate $n^{(i)}_{V,t-1} \sim \mathcal{N}\left(0,1\right)$ and $n^{(i)}_{\theta,t-1}\sim \mathcal{N}\left(0,1\right)$ & & \\
 & \textbf{ii.} simulate $n^{(i)}_{V} \sim \mathcal{N}\left(0,1\right)$ and $n^{(i)}_{\theta} \sim \mathcal{N}\left(0,1\right)$ and square the results & & \\ 
 & to obtain $\left(n^{(i)}_{V}\right)^2$ and $\left(n^{(i)}_{\theta}\right)^2$ & & \\
 & \textbf{iii.} simulate $\zeta^{(i)}_{1},\zeta^{(i)}_{2},\mu^{(i)}_{1,p},\mu^{(i)}_{2,p}$ each independently from $\mathcal{N}\left(0,1\right)$. & & \\ 
 & {\it For $r=1,\ldots,p$ }& & \\ 
 & \textbf{iv.} simulate $\psi^{(i)}_{1,r},\psi^{(i)}_{2,r},\nu^{(i)}_{1,r},\nu^{(i)}_{2,r}$ each independently from $\mathcal{N}\left(0,1\right)$. & & \\
\multicolumn{4}{l}{{\textbf 3. Evaluate:}}\\
 & {\it For $i=1,\ldots,N$ }& & \\ 
 &\textbf{i.} evaluate $J_{2,1}^p{\theta}$ and $J_{1,2}^p(V)$ using equation \ref{eqn:Jmix} and simulated terms from stage 2. & & \\
 &\textbf{ii.} evaluate Milstein approximations to obtain new samples $\theta_t^{(i)}$ and $V_t^{(i)}$ given $\theta_{t-1}^{(i)}$ and $V_{t-1}^{(i)}$  & & \\
\hline
\end{tabular}
\caption{\small Generation from the 2-d Milstein strong stochastic Taylor scheme, truncated at p-th order. }
\end{table}

\newpage
\section{Appendix 4: Adaptive Metropolis within Rao-Blackwellised Particle MCMC Algorithm}
\label{IRAlgs}
{\small{
\begin{algorithm}
\dontprintsemicolon \KwIn{ Initial Markov chain state
$\left(\bm{\Phi}^{(0)}, \chi_{1:t}^{(0)}, \xi_{1:t}^{(0)},
\theta_{1:t}^{(0)}, V_{1:t}^{(0)}\right)$. }
\KwOut{J Markov chain samples $\{\bm{\Phi}^{(j)},\chi_{1:t}^{(j)},
\xi_{1:t}^{(j)}, \theta_{1:t}^{(j)}, V_{1:t}^{(j)} \}_{j=1:J} \sim
p\left( \bm{\Phi},\chi_{1:t}, \xi_{1:t}, \theta_{1:t},
V_{1:t}|\bm{F_{1:t,T}}\right)$.}
\Begin{
\begin{enumerate}
\item Set initial state $\left(\bm{\Phi}^{(0)}, \chi_{1:t}^{(0)}, \xi_{1:t}^{(0)},
\theta_{1:t}^{(0)}, V_{1:t}^{(0)}\right)$ deterministically or by
sampling the priors.\;
\end{enumerate}
\Repeat{j=J}{
\begin{enumerate}
\item[2.] Sample $\bm{\Phi}$, given $\bm{\Phi}^{(j-1)}$, from an adaptive multivariate Gaussian Metropolis proposal, to obtain proposal $\bm{\Phi}'$\;
\item[3.] Sample $\chi_{1:t},\xi_{1:t},\theta_{1:t},V_{1:t}$ via a combination of Rao-Blackwellizing Kalman Filter and
SIR particle filtering, conditional on $\bm{\Phi}'$ to obtain proposal
$\left(\bm{\Phi}',\chi'_{1:t},\xi'_{1:t},\theta'_{1:t},V'_{1:t}\right)$\;
\item[4.] Accept proposal $\left(\bm{\Phi}',\chi'_{1:t},\xi'_{1:t},\theta'_{1:t},V'_{1:t}\right)$ with Metropolis-Hastings acceptance probability
given by
\begin{align*}
&\alpha\left(\left(\bm{\Phi}^{(j-1)},\chi_{1:t}^{(j-1)},\xi_{1:t}^{(j-1)},\theta_{1:t}^{(j-1)},V_{1:t}^{(j-1)}\right),\left(\bm{\Phi}',\chi'_{1:t},\xi'_{1:t},\theta'_{1:t},V'_{1:t}\right)\right)\\
&= \min
\left(1,\frac{\widehat{p}\left(\bm{F_{1:t,T}}|\bm{\Phi}'\right)p\left(\bm{\Phi}'\right)q\left(\bm{\Phi}',\bm{\Phi}^{(j-1)}\right)}
 {\widehat{p}\left(\bm{F_{1:t,T}}|\bm{\Phi}^{(j-1)}\right)p\left(\bm{\Phi}^{(j-1)}\right)q\left(\bm{\Phi}^{(j-1)},\bm{\Phi}'\right)} \right)
\end{align*}
\; and increment $j = j + 1$.
\end{enumerate}
}} \caption{Adaptive Metropolis within Rao-Blackwellised Particle
MCMC\label{IR}}
\end{algorithm}
}}

\noindent \rule{\textwidth}{1pt}\\
\textbf{Step 2: Sample the static model parameters.} \\
{\small{
\eIf(\tcc*[f]{perform an adaptive random walk move}){$u_1 \geq
w_1$}{
\begin{enumerate}
\item[2(i).] Estimate $\Sigma_j$ the empirical covariance, at iteration $j$ of the Markov chain, for $\bm{\Phi}$ using samples $\{\bm{\Phi}^{(i)}\}_{i=1:j}$.\;
\item[2(ii).] Sample proposal $\bm{\Phi}' \sim N\left(\bm{\Phi};\bm{\Phi}^{(j-1)},\frac{\left(2.38\right)^2}{d}\Sigma_j\right)$.\;
\end{enumerate}
}(\tcc*[f]{perform a non-adaptive random walk move}) {
\begin{enumerate}
\item[2(i).] Sample proposal $\bm{\Phi}' \sim N\left(\bm{\Phi};\bm{\Phi}^{(j-1)},\frac{\left(0.1\right)^2}{d}I_{d,d}\right)$.\;
\end{enumerate}
} }} \noindent \rule{\textwidth}{1pt}

{\small{
\begin{algorithm}
\begin{enumerate}
\item[3(i).]{Initialise the SIR particle filter for $\bm{X}_0=(\theta_0,V_0)$ with $N$ particles $\{\left(\theta^{(i)}_0,V^{(i)}_0\right)\}_{i=1:N}$}
\item[3(ii).]{Initialise the Kalman filter mean and covariance for $\chi^{(i)}_0,\xi^{(i)}_0$ denoted
by $\bm{\mu}_t^{(i)}=\left(\mu_{\chi}^{(i)}(0),\mu_{\xi}^{(i)}(0)\right)$ and
$\Sigma_t^{(i)} = \Sigma^{(i)}_{\left(\chi,\xi\right)(0)}$}
\end{enumerate}
\Repeat{t=T}{ 
\begin{enumerate}
\item[3(iii).]{Perform Kalman filter update for the $N$ particles at iteration $t-1$, to update $\bm{\mu}_{t-1}^{(i)}$ and $\Sigma_{t-1}^{(i)}$ \\ to $\bm{\mu}_{t}^{(i)}$ and $\Sigma_{t}^{(i)}$ according to recursions in (Section 6.3)}
\item[3(iv).]{Perform Mutation of the $N$ particles at iteration $t-1$ to obtain new particles at iteration $t$ via\\
prior, $\bm{X}_t^{(i)} \sim \mathcal{N}\left(\bm{x}_t;
f_{t}\left(\bm{x}_{t-1},\bm{\theta}'\right),
\sigma_{\epsilon}^2\right)$. Append the new state at time $t$ to
the previous particle geneology to obtain $\bm{X}_{0:t}^{(i)}$ for $i =
1,\ldots,N$}
\item[3(v).]{Evaluate the importance sampling weight for the $N$ particles as, $i$-th weight is $\widetilde{W}^{(i)}_{t} \propto W^{(i)}_{t-1}p(y_{t}|\bm{x}^{(i)}_t,\bm{\theta}')$.}
\end{enumerate}
}
\begin{enumerate}
\item[3(vi).]{Normalise the importance sampling weights $W^{(i)}_{t}=\frac{\widetilde{W}^{(i)}_{t}}{\sum_{i=1}^{N}\widetilde{W}^{(i)}_{t}}$}
\item[3(vii).]{If the Effective Sample size is less than 80\% then resample the particles at time $t$ using stratified resampling
based on the empirical distribution constructed from the
importance weights to obtain new particles with equal weight.}
\end{enumerate}
\begin{enumerate}
\item[3(viii).]{Sample one proposed path trajectory from the particle
estimated proposal distribution given by $X'_{1:T} \sim
\widehat{p}\left(x_{1:T}|y_{1:T},\bm{\theta}'\right)
=\sum_{i=1}^{N}W_{1:T}^{(i)}\delta_{x_{1:T}^{(i)}}\left(x_{1:T}\right)$}
\item[3(ix).]{Evaluate the marginal likelihood given by $\widehat{p}\left(y_{1:T}|\bm{\theta}'\right)$.}
\end{enumerate}
\caption{Step 3: Particle Filter Proposal - Conditional on the sampled
static model parameters $\bm{\theta}'$, run an SIR particle filter
with N particles}
\end{algorithm}
}}

\end{document}